\documentclass[11pt]{article}

\usepackage[T1]{fontenc}
\usepackage[cp1251]{inputenc}
\usepackage{textcomp}
\usepackage[russian,english]{babel}
\usepackage[centertags]{amsmath}
\usepackage[mediummath]{nccmath}
\usepackage{amsfonts}
\usepackage{amssymb}
\usepackage[pdftex,driverfallback=dvipdfm]{hyperref}%hypertex
\usepackage{graphicx}
\usepackage{booktabs,array}
\usepackage[numbers,sort&compress]{natbib}% упорядочивает ссылки

\usepackage{paperinitial}% Установка параметров страницы

\paperinitialization{15mm}{15mm}{15mm}{15mm}{2pt}{10pt}

\DeclareMathOperator{\im}{Im}

\DeclareMathOperator{\rot}{rot}
\DeclareMathOperator{\Sp}{Sp}
\DeclareMathOperator{\Li}{Li}

\newcommand{\lan}{\langle}
\newcommand{\ran}{\rangle}
\newcommand{\bs}{\boldsymbol}

\newcommand{\e}{\varepsilon}
\newcommand{\vf}{\varphi}
\newcommand{\vk}{\varkappa}
\newcommand{\s}{\sigma}

\newcommand{\al}{\alpha}
\newcommand{\be}{\beta}
\newcommand{\ga}{\gamma}
\newcommand{\Ga}{\Gamma}
\newcommand{\de}{\delta}
\newcommand{\De}{\Delta}

\newcommand{\la}{\lambda}

\newcommand{\spx}{\mathbf{x}}
\newcommand{\spy}{\mathbf{y}}

\newcommand{\spp}{\mathbf{p}}
\newcommand{\spk}{\mathbf{k}}
\newcommand{\spe}{\mathbf{e}}
\newcommand{\spA}{\mathbf{A}}

\newcommand{\sle}{\selectlanguage{english}}

\newcolumntype{M}[1]{>{\centering\arraybackslash}m{#1}}

\begin{document}
\allowdisplaybreaks[4]% позволяет переносить многострочные формулы
%\frenchspacing% уменьшение пробелов после запятых
\setlength{\unitlength}{1pt}% устанавливает единицу длины в окружении picture
\sle

\title{\Large\textbf{Surface photoelectric effect by\\ twisted photons as a source of twisted electrons}}
% Surface photoelectric effect by twisted photons
% Surface photoelectric effect by twisted photons as the source of twisted electrons

\date{}

\author{P.O. Kazinski${}^{1)}$\thanks{E-mail: \texttt{kpo@phys.tsu.ru}},\; M.V. Mokrinskiy${}^{1)}$\thanks{E-mail: \texttt{mark.mokrinskiy@bk.ru}},\;and V.A. Ryakin${}^{2)}$\thanks{E-mail: \texttt{vlad.r.a.phys@yandex.ru}}\\[0.5em]
{\normalsize ${}^{1)}$ Physics Faculty, Tomsk State University, Tomsk 634050, Russia}\\[0.5em]
{\normalsize ${}^{2)}$Mathematics and Mathematical Physics Division,}\\ {\normalsize Tomsk Polytechnic University, Tomsk 634050, Russia}
}

\maketitle

\begin{abstract}

The theory of surface photoelectric effect by twisted photons is developed. The explicit expression for the probability to record a twisted photoelectron is derived. The conditions when the surface photoelectric effect can be used as a pure source of twisted electrons are found. It is shown that the lightly doped n-InSb crystal with interface without defects at temperatures lower than $2.5$ K satisfies these conditions. The Dirac and Weyl semimetals with electron chemical potential near the top of the Dirac cone obey these conditions at temperatures lower than $60$ K and can also be employed for design of pure sources of twisted electrons by the photoelectric effect.

\end{abstract}

\section{Introduction}

The photoelectric effect is one of the phenomena whose explanation encouraged creation of quantum theory and led to introduction of quanta of light -- photons (\cite{Einstein1905}, see also \cite{Tartakovskii1940} for a review of early papers). The simplest quantum mechanical description of photoelectric effect in metals with account for the influence of the crystal boundary on the electromagnetic field of the electron was given in \cite{Mitchell1935}. Further, this theory was considerably developed in  \cite{Makinson1949,Herring1949,Wright1950,Mahan1970,Sehaich1971,Penn1972,Endriz1973,Feibelman1975,Kliewer1976,Bagchi1978,Raseev2007} where the effects of band structure \cite{Mahan1970}, of crystal interface \cite{Makinson1949,Herring1949,Wright1950,Endriz1973,Kliewer1976}, of interaction of electrons \cite{Sehaich1971,Endriz1973,Feibelman1975,Raseev2007}, and of possible surface states \cite{Penn1972,Bagchi1978,Raseev2007} were scrutinized. The multiphoton photoelectric effects was also studied \cite{Gladun1970}. On the other hand, rather recently, a considerable progress has been achieved in investigating the effects induced by quantum states of electrons, photons, and other particles with definite projection of the angular momentum on a certain axis, which are called twisted states \cite{Bliokh2017,Lloyd2017,SerboNew,New18,New19,OAMPM} (for some recent papers in high energy physics see, e.g., \cite{Bogdanov2019,Bogdanov2020,BKLb,KazSok2024,Lu2023,Lu2024,Sizykh2024,Balabanski2024,Korotchenko2024,Epp2023,Epp2024,Bu2024} and in condensed matter physics see, e.g., \cite{WCBA14,KonKruGie19,Pattan22,GrBhSeHa22,QRTKrmp22,Busch2023,KazRyak23,Matsumoto2024}). Despite the fact that several techniques to produce twisted electrons have been elaborated at present \cite{Bliokh2017,Lloyd2017,Pavlov2024,Huo2024,Ababekri2024}, there is a need in new pure sources of twisted electrons that admit a simple experimental realization.

In the present paper, we study the surface photoelectric effect where the crystal is irradiated by a twisted photon or by a beam of twisted photons. Naively, one may expect that the angular momentum bearing by the twisted photon is transmitted to the photoelectron and that allows one to generate twisted electrons with the aid of the photoelectric effect. However, it turns out that this anticipation is invalid in general. Only at low temperatures in doped semiconductors with small effective electron mass as, for example, in n-InSb or in Dirac or Weyl semimetals is the angular momentum of twisted photon almost completely transferred to the photoelectron. Only in this case can the photoelectric effect induced by twisted photons be harnessed as a pure source of twisted electrons. We shall find the restrictions on the parameters of a crystal, of photoelectrons, and of incident twisted photons securing this. In particular, it follows from these estimates that the photoelectric effect in ordinary metals cannot be used as a pure source of twisted electrons.

In order to provide an analytical description of the photoelectric effect by twisted photons, we employ the simplest quantum model that catches main physics of this effect and even gives quantitative predictions for the surface photoelectric effect near its threshold \cite{Brodskii1968,Brodskii1971,LevichVdovinMyamlII,BrodskiiGurevich1973}. We describe the conduction electrons in the crystal in the framework of the effective mass approximation (the envelope approximation) for a single valley \cite{Kittel63,LandLifshST2,ShklovEfr_book1984}. The interaction of an electron with other electrons, with impurities, and with phonons is neglected in the leading order of perturbation theory for the given temperatures and concentrations of conduction electrons and impurities. The influence of the crystal interface on quantum dynamics of a conduction electron is taken into account in the form of a screened Coulomb interaction with the image of this electron, the usual Debye screening in the crystal being supposed. As is known \cite{Endriz1973} and as it will be seen from the formalism developed in the present paper, a poorly known electric potential in the surface layer weakly affects the angular dependence of the amplitude of photoelectric effect and, consequently, weakly affects the decomposition of a photoelectron state over the projections of angular momentum onto the normal to the crystal interface. Yet, the crystal interface is assumed to be perfectly flat, i.e., without defects. This condition is important for the surface photoelectric effect to be employed as a pure source of twisted electrons.

The paper is organized as follows. We start in Sec. \ref{Hamiltonian_sec} with formulation of the model and the main approximations. In  Sec. \ref{Probability_Sec}, we derive the general formula for the differential probability to record a photoelectron. In this section we extensively use the formalism developed in \cite{KazSol22,radet24} for evaluation of inclusive probabilities to detect particles in quantum field theory processes. Section \ref{Surface_Photoel_Sec} is devoted to calculation of the differential probability to record a twisted photoelectron ejected from the crystal by a twisted photon. Here we deduce the general formula for this probability and find the conditions when the surface photoelectric effect can be used as a pure source of twisted electrons. In Conclusion we summarize the results. In Appendix \ref{Electron_Modes_App}, we derive the modes of the electron quantum field, whereas in Appendix \ref{Photon_Modes_App}, we do the same for photons. We use the system of units such that $c=\hbar=k_B=1$ and $e^2=4\pi\al$, where $\alpha$ is the fine structure constant.

\section{Hamiltonian}\label{Hamiltonian_sec}

Let us consider the interaction of the conduction electrons in the crystal with the quantum electromagnetic field. Suppose that the crystal constitutes a plate perpendicular to the $z$ axis and is located at $z<0$. For $z>0$, there is a vacuum. The transition layer between the crystal and the vacuum has the width of order $\de\approx 1-2$ {\AA} \cite{BrodskiiGurevich1973,Brodskii1968,Bagchi1978,Endriz1973}. To describe the surface photoelectric effect caused by the incident twisted photon, we consider the simplest model of conduction electrons in the crystal in the effective mass approximation disregarding the spin degree of freedom of electrons. The presence of the electron spin is taken into account only as degeneracy multiplicity of the electronic levels. We assume that the dispersion law of conduction electrons has only one valley near the point $\spp=0$ where it approximately has a parabolic form. Then the Hamiltonian of the system reads
\begin{equation}\label{Hamiltonian}
\begin{split}
    \hat{H}&=\int d\spx\hat{\psi}^\dag(\spx)h_0\hat{\psi}(\spx)+\hat{H}_{int}+\hat{V}_{Coul}+\hat{H}_{em},\\
    h_0&=p_i\frac{\de_{ij}}{2\tilde{m}}p_j+U(z),\\
    \hat{H}_{int}&=\hat{H}_{int}^{(1)}+\hat{H}_{int}^{(2)}:=-\int d\spx\hat{\psi}^\dag(\spx) \big[\spp\hat{\spA}(\spx)\frac{e}{2\tilde{m}}+\frac{e}{2\tilde{m}}\hat{\spA}(\spx)\spp\big] \hat{\psi}(\spx)+\int d\spx \frac{e^2}{2\tilde{m}}\hat{\psi}^\dag(\spx) \hat{\spA}^2(\spx)\hat{\psi}(\spx),
\end{split}
\end{equation}
where $\tilde{m}$ is the effective mass of conduction electrons,
\begin{equation}\label{Uz}
    U(z)=
    \left\{
      \begin{array}{ll}
        -\frac{e^2 e^{-|z|/r_D}}{16\pi\e_0 z}\frac{\e_0-1}{\e_0+1}-U_0, & \hbox{for $z<-\delta_-/2$;} \\[1em]
        -\frac{e^2}{16\pi z}\frac{\e_0-1}{\e_0+1} +U_{ext}(z), & \hbox{for $z>\delta_+/2$,}
      \end{array}
    \right.
\end{equation}
and $\e_0$ is the static dielectric constant of the crystal, $r_D$ is the Debye radius, $\de_\pm\sim \de$ characterize the width of the transition layer, $U_{ext}(z)$ is the potential energy of interaction of a photoelectron with an applied electric field, $-U_0$ is the potential energy of the bottom of conduction band with respect to the vacuum zero level, the operator $\hat{H}_{em}$ is the Hamiltonian of the free electromagnetic field. The first term in expression \eqref{Uz} describes the energy of interaction of the electron with its image \cite{LandLifshECM}. For a metallic plate, $\e_0\rightarrow-\infty$. In the transition layer, $z\in[-\de_-/2,\de_+/2]$, the interaction potential $U(z)$ has a rather complicated form. However, in virtue of the fact that we will consider the photons with energies of  order $1-10$ eV and the photoelectrons with momenta not exceeding several keVs (the kinetic energy of photoelectrons is of order of or smaller than several eVs), the form of the interaction potential in the transition layer is irrelevant \cite{Brodskii1968,Brodskii1971,LevichVdovinMyamlII,BrodskiiGurevich1973}. The electron wave functions can be joined at the boundaries of the transition layer because they do not change appreciably when $z$ varies from $-\de_-/2$ to $\de_+/2$. The effective electron mass, $\tilde{m}$, depends on $z$. We suppose that for $z<-\de_-/2$ the effective mass $\tilde{m}=m_*=const$, whereas for $z>\de_+/2$ the effective electron mass coincides with the electron mass in vacuum, $\tilde{m}=m_0$.

The operator of a screened Coulomb interaction between electrons takes the standard form
\begin{equation}
    \hat{V}_{Coul}=-\frac{e^2}{2}\int d\spx d\spy \hat{\psi}^\dag(\spx)\hat{\psi}(\spx)\De^{-1}(\spx,\spy)
    \hat{\psi}^\dag(\spy)\hat{\psi}(\spy),
\end{equation}
where $\De^{-1}(\spx,\spy)$ is the potential of a unit point charge located at the point $\spy$ in the presence of the plate with static dielectric constant $\e_0$ \cite{LandLifshECM}. The field operators are written as
\begin{equation}\label{field_operators}
\begin{split}
    \hat{\psi}(\spx)&=\sum_{\al} \frac{\psi_\al(\spx)}{\sqrt{V}} u_0(\spx)\hat{a}_\al,\qquad \hat{\psi}^\dag(\spx)=\sum_{\al} \frac{\psi^*_\al(\spx)}{\sqrt{V}} u^*_0(\spx) \hat{a}^\dag_\al,\\
    \hat{\spA}(\spx)&=\sum_\ga\Big[\frac{\bs\phi_\ga(\spx)}{\sqrt{2k_{0\ga}V}}\hat{c}_\ga +\frac{\bs\phi^*_\ga(\spx)}{\sqrt{2k_{0\ga}V}}\hat{c}^\dag_\ga\Big],
\end{split}
\end{equation}
where $\hat{a}^\dag_\al$ and $\hat{a}_\al$ are the creation and annihilation operators for electrons, whereas $\hat{c}^\dag_\ga$ and $\hat{c}_\ga$ are the creation and annihilation operators for photons, $V$ is the normalization volume, and $k_{0\ga}$ is the energy of a photon with quantum numbers $\ga$.

The electron mode functions (the envelopes) $\psi_\al(\spx)$ constitute the complete orthonormal basis of eigenfunctions of the operator $h_0$ and are enumerated by the quantum numbers $\al$:
\begin{equation}\label{Schroed_eqn}
    h_0\psi_\al=E_\al\psi_\al.
\end{equation}
The function $u_0(\spx)$ is the Bloch function of the electron in the periodic crystal field near the edge of the conduction band \cite{Kittel63,LandLifshST2,ShklovEfr_book1984}. We suppose that this edge is located in a small vicinity of the point $\spp=0$. In this case, $u_0(\spx)$ is a periodic function of $\spx$ and can be developed as the Fourier series with respect to the vectors of the reciprocal lattice. The typical magnitude of the reciprocal lattice vector is of order $10$ keV. As long as we will investigate the photoelectric effect for the photon and electron momenta belonging to the ranges discussed above, only the constant contribution can be retained in the Fourier expansion of  $u_0(\spx)$. Therefore, we take $u_0(\spx)=1$ and match the wave function envelope, $\psi_\al(\spx)$, at the crystal boundary. This is the standard approximation in description of the surface photoelectric effect \cite{Brodskii1968,Sehaich1971,Endriz1973}. The lower is the energy of photoelectrons, i.e., the photoelectric effect is studied near the threshold, and the smaller is the effective electron mass in the crystal, the better this approximation works. The decrease of the effective electron mass diminishes the magnitude of the electron momentum in the crystal,
\begin{equation}\label{electron_mom_in_cr}
    p=\sqrt{2m_*T_k},
\end{equation}
for a given kinetic energy $T_k$. For the estimates in what follows, we provide here the values of the parameters appearing in describing the dynamics of electrons in the crystal in the considered model \cite{Kittel63,Ioffe_Inst}. The effective mass of conduction electrons in InSb: $m_*=0.014 m_0$; in Cu: $m_*=m_0$. The static dielectric constant in InSb: $\e_0=16.8$; in Cu: $\e_0=-\infty$. For InSb, the energy $U_0=4.59$ eV; for Cu: $U_0=11.36$ eV. The Fermi energy for Cu: $E_F=7.0$ eV. The explicit expressions for the mode functions in the case $U_{ext}(z)=0$ are given in Appendix \ref{Electron_Modes_App}.

The photon mode functions $\bs\phi_\ga(\spx)$ constitutes the complete orthonormal set of solutions to the Maxwell equations (see, e.g., \cite{BKL5,BKKL21}),
\begin{equation}
    (k_0^2\e(k_0)\de_{ij}-\rot^2_{ij})\phi_{j\ga}=0,
\end{equation}
in the Coulomb gauge. For $z>0$, the dielectric permittivity $\e(k_0)=1$ (the presence of transition layer is neglected). For $z<0$, the dielectric permittivity is $\e(k_0)$. For a metallic plate, the dielectric permittivity can be taken in the form
\begin{equation}\label{plasma_permit}
    \e(k_0)=1-\omega_p^2/k_0^2,
\end{equation}
where $\omega_p$ is the plasma frequency. The experimental values of $\e(k_0)$ for different materials can be found, for example, in \cite{RefrIndex}. At the crystal boundary, the photon mode functions obey the standard matching conditions
\begin{equation}\label{joint_conds}
    [\bs\phi_{\ga_\perp}]_{z=0}=0,\qquad [\rot\bs\phi_{\ga_\perp}]_{z=0}=0,
\end{equation}
where the index $\perp$ denotes the vector components perpendicular to the normal to the interface and the square brackets mean the jump of the function in passing through the surface of discontinuity of the dielectric permittivity. The explicit expressions for the mode functions of the quantum electromagnetic field are presented in Appendix \ref{Photon_Modes_App}.

\section{Probability to record a photoelectron}\label{Probability_Sec}

The probability to record a photoelectron reads \cite{KazSol22,radet24}
\begin{equation}\label{inclus_prob}
    P=\Sp(\hat{\Pi}_{D}\hat{S}\hat{R}\hat{S}^\dag),
\end{equation}
where the operators are given in the interaction representation and the self-adjoint projection operator,
\begin{equation}
    \hat{\Pi}_D=(1-:\exp(-\hat{a}^\dag_\al D_{\al\be} \hat{a}_\be):)\otimes \hat{1}_{ph},
\end{equation}
projects to the states of the electrons that contain at least one state singled out by the self-adjoint projector $D_{\al\be}$. The one-particle projector $D_{\al\be}$ specifies the states of recorded photoelectrons. We assume that the operator $D_{\al\be}$ is diagonal in the energy basis and the energy of recorded electrons $E_{\bar{\al}}>0$. The density operator of the initial state of the system is taken in the form
\begin{equation}
    \hat{R}=\hat{R}_e\otimes \hat{R}_{ph},
\end{equation}
where the operator,
\begin{equation}
    \hat{R}_{ph}=|d\ran\lan\bar{d}|e^{-\bar{d}d},
\end{equation}
defines the coherent state of photons with the complex amplitude $d_\ga$. As for the initial state of electrons, it is taken in the form of the Fermi-Dirac distribution so that (see formula (140) of \cite{KazSol22})
\begin{equation}
    \lan\bar{a}|\hat{R}_e|a\ran=\exp(\bar{a}e^{-\be_T\tilde{\e}}a)/Z,\qquad\ln Z=\Sp\ln(1+e^{-\be_T\tilde{\e}}),
\end{equation}
where $\tilde{\e}_{\al\be}=(E_\al-\mu)\de_{\al\be}$ (no summation over $\al$), $\be_T$ is the reciprocal temperature, $\mu$ is the chemical potential with respect to the zero vacuum level, and $E_\al$ is the energy of a one-particle electron state. Henceforth, we employ the Bargmann-Fock representation (see for details, e.g., \cite{KazSol22,ippccb}).

In order to simplify further calculations we assume that thermionic emission can be neglected. Formally, this implies the fulfilment of the approximate equalities
\begin{equation}\label{Pi_D_orthogon}
    \hat{\Pi}_D\hat{R}_e\approx \hat{R}_e\hat{\Pi}_D\approx0,
\end{equation}
i.e., the probability to record the electron with positive energy in the initial Fermi-Dirac distribution is approximately zero. All the contributions to the $S$-matrix up to the second order in the coupling constant are listed in the paper \cite{KazSol22}. If the condition \eqref{Pi_D_orthogon} is met, only one nonzero contribution to \eqref{inclus_prob} is left in the second order of perturbation theory
\begin{equation}
    P=\Sp(\hat{\Pi}_D\hat{V}\hat{R}\hat{V}^\dag)=-\Sp(\hat{\Pi}_D\hat{V}\hat{R}\hat{V}),
\end{equation}
where
\begin{equation}
    \hat{V}=V^{\bar{\ga}}_{\bar{\al}\al}\hat{a}^\dag_{\bar{\al}}\hat{a}_\al\hat{c}^\dag_{\bar{\ga}} -(V^\dag)^{\ga}_{\bar{\al}\al}\hat{a}^\dag_{\bar{\al}}\hat{a}_\al\hat{c}_{\ga}.
\end{equation}
Whence we obtain
\begin{equation}\label{inclus_prob1}
\begin{split}
    P=&\Sp(\hat{R}_e \hat{a}^\dag_{\bar{\al}} \hat{a}_{\al} \hat{\Pi}_D \hat{a}^\dag_{\bar{\be}} \hat{a}_{\be}) \big[V^{\bar{\ga}}_{\bar{\al}\al} (V^\dag)^{\ga}_{\bar{\be}\be} \Sp(\hat{R}_{ph}\hat{c}^\dag_{\bar{\ga}} \hat{c}_{\ga}) +(V^\dag)^{\ga}_{\bar{\al}\al} V^{\bar{\ga}}_{\bar{\be}\be}  \Sp(\hat{R}_{ph} \hat{c}_{\ga} \hat{c}^\dag_{\bar{\ga}})-\\
    &-V^{\bar{\ga}'}_{\bar{\al}\al} V^{\bar{\ga}}_{\bar{\be}\be}  \Sp(\hat{R}_{ph} \hat{c}^\dag_{\bar{\ga}'} \hat{c}^\dag_{\bar{\ga}}) -(V^\dag)^{\ga'}_{\bar{\al}\al} (V^\dag)^{\ga}_{\bar{\be}\be}  \Sp(\hat{R}_{ph} \hat{c}_{\ga} \hat{c}_{\ga}) \big].
\end{split}
\end{equation}
The trace over the fermionic degrees of freedom is readily evaluated (see formula (209) of \cite{radet24}),
\begin{equation}\label{trace}
    \Sp(\hat{R}_e \hat{a}^\dag_{\bar{\al}} \hat{a}_{\al} \hat{\Pi}_D \hat{a}^\dag_{\bar{\be}} \hat{a}_{\be})=\de_{\al\bar{\be}} (\rho^{(1)}-\rho^{(1)}_{\tilde{D}})_{\be\bar{\al}} +D_{\al\bar{\be}} (\rho^{(1)}_{\tilde{D}})_{\be\bar{\al}} -\rho^{(2)}_{\al\be|\bar{\al}\bar{\be}} +\tilde{D}_{\al\al_1}\rho^{(2)}_{\al_1\be|\bar{\al}\bar{\be}_1}\tilde{D}_{\bar{\be}_1\bar{\be}}.
\end{equation}
As the convolution of the projector $D_{\al\bar{\be}}$ with the $N$-particle density matrix with respect to any index is approximately zero, one can replace $\tilde{D}_{\al\bar{\be}}\rightarrow\de_{\al\bar{\be}}$ on the right-hand side of \eqref{trace}. Therefore,
\begin{equation}\label{ferm_trace}
    \Sp(\hat{R}_e \hat{a}^\dag_{\bar{\al}} \hat{a}_{\al} \hat{\Pi}_D \hat{a}^\dag_{\bar{\be}} \hat{a}_{\be})\approx D_{\al\bar{\be}} \rho^{(1)}_{\be\bar{\al}}=D_{\al\bar{\be}}n^f_{\be\bar{\al}},
\end{equation}
where $n^f_{\be\bar{\al}}=[\exp(\be_T\tilde{\e})+1]^{-1}_{\be\bar{\al}}$.

Now we take into account the energy conservation law in the transition amplitude $(V^\dag)^\ga_{\bar{\be}\be}$. This transition amplitude corresponds to the process
\begin{equation}
    e_\be+k_{0\ga}\rightarrow e_{\bar{\be}}.
\end{equation}
The amplitude $V^{\bar{\ga}}_{\bar{\al}\al}$ describes the process
\begin{equation}
    e_\al\rightarrow e_{\bar{\al}}+k_{0\bar{\ga}}.
\end{equation}
It follows from expression \eqref{ferm_trace} that the quantum numbers $\al$ and $\bar{\be}$ correspond to the positive energies of the recorded photoelectron, while the quantum numbers $\bar{\al}$ and $\be$ are associated with the negative energies of the electron in the crystal. Consequently, the energy conservation law is satisfied only for the first term in the square brackets in \eqref{inclus_prob1}. As a result, evaluating the trace over the photonic degrees of freedom, we arrive at
\begin{equation}\label{probab_gen}
    P=D_{\al\bar{\be}} d_\ga (V^\dag)^{\ga}_{\bar{\be}\be} n^f_{\be\bar{\al}} V^{\bar{\ga}}_{\bar{\al}\al}\bar{d}_{\bar{\ga}}.
\end{equation}
This result is known \cite{Mitchell1935} and is anticipated on physical grounds: the one-particle probability of photoemission of an electron is averaged with the aid of the initial Fermi-Dirac distribution.

Bearing in mind the form of the interaction Hamiltonian in the leading order of perturbation theory, we come to
\begin{equation}\label{V_ampl}
\begin{split}
    V^{\bar{\ga}}_{\bar{\al}\al} =&\int \frac{dt d\spx ie}{2\tilde{m} V\sqrt{2Vk_0^{\bar{\ga}}}} e^{i(E_{\bar{\al}\al} +k_0^{\bar{\ga}})t-i(\spp^\perp_{\bar{\al}\al} +\spk^{\bar{\ga}}_\perp)\spx_\perp}\big[(\spp^\perp_{\bar{\al}} +\spp^\perp_{\al})_i (\phi^*_{\bar{\ga}})^\perp_i \psi^*_{\bar{\al}}\psi_\al +i(\phi^*_{\bar{\ga}})_3 (\psi'^*_{\bar{\al}}\psi_\al -\psi^*_{\bar{\al}}\psi'_\al) \big]=\\
    =&\;ie(2\pi)^3\frac{\de(E_{\bar{\al}\al} +k_0^{\bar{\ga}}) \de(\spp^\perp_{\bar{\al}\al} +\spk^{\bar{\ga}}_\perp)}{V\sqrt{2Vk_0^{\bar{\ga}}}} w^{\bar{\ga}}_{\bar{\al}\al},
\end{split}
\end{equation}
where $E_{\bar{\al}\al}:=E_{\bar{\al}}-E_\al$, $\spp_{\bar{\al}\al}=\spp_{\bar{\al}}-\spp_\al$ and
\begin{equation}\label{wbg_baa}
    w^{\bar{\ga}}_{\bar{\al}\al}:=\int_{-\infty}^\infty \frac{dz}{2\tilde{m}} \big[(2\spp^\perp_{\al}-\spk^\perp_{\bar{\ga}})_i (\phi^*_{\bar{\ga}})^\perp_i \psi^*_{\bar{\al}}\psi_\al +i(\phi^*_{\bar{\ga}})_3 (\psi'^*_{\bar{\al}}\psi_\al -\psi^*_{\bar{\al}}\psi'_\al) \big].
\end{equation}
All the mode functions in this expression depend on $z$ and are presented in Appendices \ref{Electron_Modes_App} and \ref{Photon_Modes_App}. Using the property of the Schr\"{o}dinger equation,
\begin{equation}
    \big[\frac{1}{2\tilde{m}}(\psi^*_{\bar{\al}}\psi'_\al -\psi'^*_{\bar{\al}}\psi_\al)\big]'=E_{\bar{\al}\al}\psi^*_{\bar{\al}} \psi_\al,
\end{equation}
and integrating by parts in the second term in \eqref{wbg_baa}, we rewrite \eqref{wbg_baa} as
\begin{equation}\label{wbg_baa1}
    w^{\bar{\ga}}_{\bar{\al}\al}:=\int_{-\infty}^\infty \frac{dz}{2\tilde{m}} \psi^*_{\bar{\al}}\psi_\al \big[(2\spp^\perp_{\al}-\spk^\perp_{\bar{\ga}})_i (\phi^*_{\bar{\ga}})^\perp_i  -2\tilde{m}i k_0^{\bar{\ga}} (\Phi^*_{\bar{\ga}})_3 \big],
\end{equation}
where
\begin{equation}
    (\Phi_{\bar{\ga}})_3(z):=\int_0^z dz'(\phi_{\bar{\ga}})_3(z').
\end{equation}
The integrand of \eqref{wbg_baa1} is finite for any $z$ as against the integrand of \eqref{wbg_baa}. Moreover, the last term in the square brackets in \eqref{wbg_baa1} is small in the vicinity of the point $z=0$ where the electron mode functions are poorly known. Further, one needs to substitute expression \eqref{V_ampl} into \eqref{probab_gen}, to specify the projector defining the measurement of the electron state, and to introduce the initial coherent state of photons.

\section{Surface photoelectric effect induced by twisted photons}\label{Surface_Photoel_Sec}

We are aimed to find the probability that the photoelectron ejected from the crystal surface by a twisted photon has the quantum numbers $\al_0=(l_0,p_3^0,p_\perp^0)$, where $l_0$ is the projection of orbital angular momentum onto the $z$ axis, $p_3^0$ is the projection of momentum onto this axis, and $p_\perp^0$ is the absolute value of the momentum component perpendicular to the $z$ axis. The free electron states with these quantum numbers have the form \cite{Bliokh2017,Lloyd2017}
\begin{equation}
    |\al_0\ran = \frac{J_{l_0}(p_\perp^0 r)}{\sqrt{2RL p_\perp^0}} e^{i(l_0\vf +p_3^0 z)},
\end{equation}
where $R$ and $L$ are the radius and the length of a large cylinder where these states are normalized to unity, $r=\sqrt{x^2+y^2}$, $\vf=\arg(x+iy)$. These states form a complete orthonormalized set
\begin{equation}
    \sum_{\al_0}|\al_0\ran\lan\al_0|=\de(\spx-\spy),\qquad \lan\al_0|\be_0\ran=\de_{\al_0\be_0},
\end{equation}
where
\begin{equation}
    \sum_{\al_0}=\sum_{l_0=-\infty}^\infty \int_0^\infty \frac{R dp_\perp^0}{\pi}\int_{-\infty}^\infty \frac{L dp_3^0}{2\pi}.
\end{equation}
Then the projection operator describing the measurement of the electron in the state with quantum numbers $\al_0$ becomes
\begin{equation}
    D_{\al\bar{\be}}=\lan\al|\al_0\ran\lan\al_0|\bar{\be}\ran d\al_0=(2\pi)^2e^{il_0(\vf_p^\al-\vf_p^{\bar{\be}})}\de(p_3^0-p_3^\al) \de(p_\perp^0-p_\perp^\al) \de(p_3^0-p_3^{\bar{\be}}) \de(p_\perp^0-p_\perp^{\bar{\be}})\frac{dp^0_\perp dp^0_3}{Vp^0_\perp},
\end{equation}
where $|\al\ran$ and $|\bar{\be}\ran$ are the plane-wave states of a free electron normalized to unity in a box of a sufficiently large volume $V$.

We choose the complex amplitude of the initial coherent state of photons in the form
\begin{equation}
    d_\ga=\sqrt{\frac{(2\pi)^3}{V}} d_{s_\ga}(\spk^\ga),\qquad d_{s_\ga}(\spk^\ga)= \de_{ss_\ga} g(k_\perp^\ga,k_3^\ga) e^{im\vf^\ga_k},
\end{equation}
where $m\in \mathbb{Z}$ is the projection of total angular momentum of photons in a coherent state onto the $z$ axis, $s=\pm1$ is the helicity of photons, $\vf^{\ga}_k=\arg(k^\ga_1+ik^\ga_2)$, and $k^\ga_3=-\sqrt{(k_0^\ga)^2-(k_\perp^\ga)^2}$. For example, for the Laguerre-Gauss states
\begin{equation}\label{GL_modes}
    g(k_\perp^\ga,k_3^\ga)=\frac{C_{ph}}{\sqrt{(2\pi)^{3/2}\s_\parallel\sigma^2_{\perp}}}   \sqrt{\frac{n!}{(n+|m|)!}} \Big({\frac{k^\ga_{\perp}}{\sqrt{2}\sigma_{\perp}}}\Big)^{|m|} L_{n}^{|m|}\Big({\frac{(k_{\perp}^\ga)^{2}}{2\sigma_{\perp}^{2}}}\Big)e^{-\frac{(k^\ga_{\perp})^2}{4\sigma_{\perp}^2}} e^{-\frac{(k_3^\ga-k_3)^2}{4\s^2_\parallel}},
\end{equation}
where $L_{n}^{|m|}(x)$ are the generalized Laguerre polynomials, $n=\overline{0,\infty}$ is the radial quantum number, and $C_{ph}$ is the normalization constant. Setting $C_{ph}=1$, we have
\begin{equation}
    \int d\spk^\ga |g(k_\perp^\ga,k_3^\ga)|^2=1.
\end{equation}
Notice that for the function $d_{s_\ga}(\spk^\ga)$ to be infinitely smooth, the function $g(k_\perp^\ga,k_3^\ga)$ has to be of the form $(k^\ga_{\perp})^{|m|} f(k_\perp^\ga,k_3^\ga)$, where $f(k_\perp^\ga,k_3^\ga)$ is an infinitely smooth function.

Due to the fact that in the basis of states we consider
\begin{equation}
    n^f_{\be\bar{\al}}=n^f_{\be\be}\de_{\be\bar{\al}}\quad\text{(no summation over $\be$)},
\end{equation}
the differential probability to detect a twisted electron \eqref{probab_gen} takes the form
\begin{equation}\label{probab_diff}
    \frac{dP(\al_0)}{dp_\perp^0 dp_3^0}=\frac{e^2}{(2\pi)^4}  \int d\spp^\be_\perp \Big[ \int_0^\infty  d\bar{p}^\be_3\theta (-E_\be) +\int_{-\infty}^\infty dp^\be_3 \Big] \frac{p_\perp^0}{E_{\al_0\be}}\frac{|\mathcal{A}_{\al_0\be}|^2}{e^{\be_T(E_\be-\mu)}+1},
\end{equation}
where we have taken into account the twofold spin degeneracy of the electron states and $E_\be$ is to be expressed in terms of $\bar{p}_3^\be$ or $p^\be_3$ as it is given after formula \eqref{sum_al}. The notation has also been introduced,
\begin{equation}\label{A_ampl}
\begin{split}
    \mathcal{A}_{\al_0\be}:=&  \int_0^\infty dk_\perp^{\bar{\ga}} k^{\bar{\ga}}_\perp \int_{k_0^{\bar{\ga}}>k_\perp^{\bar{\ga}}} \frac{dk_0^{\bar{\ga}} k_0^{\bar{\ga}}g^*(k_\perp^{\bar{\ga}},k_3^{\bar{\ga}}) }{\sqrt{(k_0^{\bar{\ga}})^2-(k^{\bar{\ga}}_\perp)^2}} \de(E_{\al_0\be} -k_0^{\bar{\ga}}) \int_0^{2\pi} d\vf_p^\al d\vf^{\bar{\ga}}_k \de(\spp^\perp_{\al\be}-\spk_\perp^{\bar{\ga}}) e^{il_0\vf^\al_p -im\vf^{\bar{\ga}}_k} w^{\bar{\ga}}_{\be\al}\big|_{s_{\bar{\ga}}=s_0}=\\
    =&\;  E_{\al_0\be} \theta(E_{\al_0\be})\int_0^{E_{\al_0\be}}   \frac{dk_\perp^{\bar{\ga}} k^{\bar{\ga}}_\perp  g^*(k_\perp^{\bar{\ga}},k_3^{\bar{\ga}}) }{\sqrt{E_{\al_0\be}^2-(k^{\bar{\ga}}_\perp)^2}}  \int_0^{2\pi} d\vf_p^\al d\vf^{\bar{\ga}}_k \de(\spp^\perp_{\al\be}-\spk_\perp^{\bar{\ga}}) e^{il_0\vf^\al_p -im\vf^{\bar{\ga}}_k} w^{\bar{\ga}}_{\be\al}\big|_{s_{\bar{\ga}}=s_0,k_0^{\bar{\ga}}=E_{\al_0\be}},
\end{split}
\end{equation}
where $p_\perp^\al=p_\perp^{0}$, $p_3^\al=p_3^{0}$, and one should put $k_3^{\bar{\ga}}=-\sqrt{E_{\al_0\be}^2-(k^{\bar{\ga}}_\perp)^2}$ in the function $g^*(k_\perp^{\bar{\ga}},k_3^{\bar{\ga}})$ .

Let us first perform the integrals with respect to the angular variables $\vf_p^\al$ and $\vf^{\bar{\ga}}_k$. From the explicit expression \eqref{wbg_baa1} for $w^{\bar{\ga}}_{\be\al}$ and from formulas \eqref{phot_mod_func_expan} and \eqref{phot_mod_func_contr}, it is seen that
\begin{equation}\label{wbg_baa1sigma}
    w^{\bar{\ga}}_{\be\al}=\sum_{\s=-1}^1 w^{\bar{\ga}}_{\s\be\al} e^{i\s(\vf^{\bar{\ga}}_k -\vf^\al_p)},
\end{equation}
where $w^{\bar{\ga}}_{\s\be\al}$ do not depend on the azimuth angles. The explicit expressions for the functions $w^{\bar{\ga}}_{\s\be\al}$ are given in Appendix \ref{Electron_Modes_App} in formulas \eqref{w_expl_s} and \eqref{w_expl_0}. Then
\begin{equation}\label{A_ampl1}
    \mathcal{A}_{\al_0\be}=  \theta(E_{\al_0\be}) E_{\al_0\be} \sum_{\s=-1}^1\int_0^{E_{\al_0\be}}   \frac{dk_\perp^{\bar{\ga}} k^{\bar{\ga}}_\perp  g^*(k_\perp^{\bar{\ga}},k_3^{\bar{\ga}}) }{\sqrt{E_{\al_0\be}^2-(k^{\bar{\ga}}_\perp)^2}} I_{l_0-\s,m-\s}(\vf^\be_p) w^{\bar{\ga}}_{\s\be\al_0},
\end{equation}
where
\begin{equation}\label{Ilm}
    I_{lm}(\vf^\be_p):=\int_0^{2\pi} d\vf_p^\al d\vf^{\bar{\ga}}_k \de(\spp^\perp_{\al\be}-\spk_\perp^{\bar{\ga}}) e^{il\vf^\al_p -im\vf^{\bar{\ga}}_k}.
\end{equation}
As we see, the evaluation of the integral over $\vf_p^\al$, $\vf^{\bar{\ga}}_k$ is reduced to the evaluation of the integral $I_{lm}(\vf^\be_p)$.

The two-dimensional delta function appearing in the integrand of \eqref{Ilm} is invariant under simultaneous rotation of the vectors $\spp_\perp^\al$, $\spp_\perp^\be$, and $\spk_\perp^{\bar{\ga}}$ by the same angle. The integrand is obviously a periodic function of $\vf_p^\al$ and $\vf^{\bar{\ga}}_k$. Therefore, changing the integration variables $\vf_p^\al\rightarrow\vf_p^\al+\vf^\be_p$, $\vf^{\bar{\ga}}_k\rightarrow\vf^{\bar{\ga}}_k +\vf^\be_p$ and employing the abovementioned invariance of the delta function, we come to
\begin{equation}\label{Ilm_phi}
    I_{lm}(\vf^\be_p)=e^{i(l-m)\vf^\be_p} I_{lm}(0),
\end{equation}
where the vector $\spp^\be_\perp$ is directed along the $x$ axis in $I_{lm}(0)$. Then the amplitude \eqref{A_ampl1} is written as
\begin{equation}\label{A_ampl2}
    \mathcal{A}_{\al_0\be}=  e^{i(l_0-m)\vf^\be_p} \theta(E_{\al_0\be}) E_{\al_0\be} \int_0^{E_{\al_0\be}}   \frac{dk_\perp^{\bar{\ga}} k^{\bar{\ga}}_\perp  g^*(k_\perp^{\bar{\ga}},k_3^{\bar{\ga}}) }{\sqrt{E_{\al_0\be}^2-(k^{\bar{\ga}}_\perp)^2}} \sum_{\s=-1}^1 I_{l_0-\s,m-\s}(0) w^{\bar{\ga}}_{\s\be\al_0}.
\end{equation}
It also follows from \eqref{Ilm_phi} that, when $p_\perp^\be=0$,
\begin{equation}\label{Ilm_sel_rule}
    I_{lm}(\vf^\be_p)=I_{lm}(0)\sim \de_{lm}.
\end{equation}
Indeed, in this case $I_{lm}(\vf^\be_p)$ must be independent of $\vf^\be_p$, which is only possible if the relation \eqref{Ilm_sel_rule} is satisfied. As a result, for $p_\perp^\be=0$, the amplitude \eqref{A_ampl2} is proportional to $\de_{l_0m}$ and so the probability \eqref{probab_diff} is different from zero only when $l_0=m$. To put it another way, the photoelectrons are twisted in this case, i.e., they possess a definite projection of the orbital angular momentum $m$.

In the general case, $p_\perp^\be\geqslant0$, solving the equations arising from setting the argument of the two-dimensional delta function to zero with respect to $\vf_p^\al$, $\vf^{\bar{\ga}}_k$, and employing the formula for composition of the delta function with a smooth function, we deduce
\begin{equation}\label{Ilm0}
    I_{lm}(0)=
    \left\{
      \begin{array}{ll}
        \frac{2\cos(m\vf^{\bar{\ga}}-l\vf^\al)}{p_\perp^\al k_\perp^{\bar{\ga}} \sin(\vf^{\bar{\ga}}-\vf^\al)}, & \hbox{with $k_\perp^{\bar{\ga}}\in(p_\perp^\al-p_\perp^\be,p_\perp^\al+p_\perp^\be)$ and $k_\perp^{\bar{\ga}}>p_\perp^\be-p^\al_\perp$;} \\[1em]
        0, & \hbox{otherwise.}
      \end{array}
    \right.
\end{equation}
The conditions on the first line are the inequalities for a triangle with the sides $p_\perp^\al$, $p_\perp^\be$, and $k_\perp^{\bar{\ga}}$, and
\begin{equation}
    \vf^\al=\arccos \frac{(p_\perp^\al)^2+(p_\perp^\be)^2-(k^{\bar{\ga}}_\perp)^2}{2p_\perp^\al p_\perp^\be},\qquad \vf^{\bar{\ga}}=\arccos \frac{(p_\perp^\al)^2-(p_\perp^\be)^2-(k^{\bar{\ga}}_\perp)^2}{2k_\perp^{\bar{\ga}} p_\perp^\be},
\end{equation}
where $\arccos x\in[0,\pi]$ and recall that $p^\al_\perp=p^0_\perp$.

In order to account for the triangle inequalities, it is convenient to make a substitution,
\begin{equation}\label{change_var}
    k_\perp^{\bar{\ga}}=p_\perp^0 -p^\be_\perp\cos\de_{\bar{\ga}},\qquad dk_\perp^{\bar{\ga}}=p^\be_\perp\sin\de_{\bar{\ga}} d\de_{\bar{\ga}},
\end{equation}
and to change the integration variable $k_\perp^{\bar{\ga}}$ by $\de_{\bar{\ga}}\in[0,\pi]$. The following inequalities must be valid
\begin{equation}\label{de_ga_region}
    \cos^2\frac{\de_{\bar{\ga}}}{2}<\frac{p_\perp^0}{p_\perp^\be},\qquad \cos\de_{\bar{\ga}}>\frac{p_\perp^0-E_{\al_0\be}}{p_\perp^\be}.
\end{equation}
The first inequality is one of the triangle inequalities. The second inequality takes into account the upper integration limit in \eqref{A_ampl}, which, in turn, follows from the energy conservation law and the requirement $k_0^{\bar{\ga}}>0$. Further, one needs to substitute \eqref{change_var} into \eqref{A_ampl2} with account for the limits of integration over $\de_{\bar{\ga}}$ and then to substitute $\mathcal{A}_{\al_0\be}$ into the probability to record a twisted photoelectron \eqref{probab_diff}. It is obvious from expression \eqref{A_ampl2} that this substitution leads to the integrand of \eqref{probab_diff} independent of $\vf^\be_p$ and so the integration over $\vf^\be_p$ in \eqref{probab_diff} gives rise to a factor of $2\pi$.

We will not give here the resulting expression in an explicit form but consider in detail the special case of small $p_\perp^\be$. As we have seen above, a pure source of twisted photoelectrons arises in the case $p_\perp^\be=0$. We assume that the states of incident photons are described by the Laguerre-Gaussian modes \eqref{GL_modes} with good accuracy. Besides, we suppose that for the majority of conduction electrons in the crystal under consideration the following estimates hold
\begin{equation}\label{est_for_pure}
    p_\perp^\be\ll p_\perp^0,\qquad p_\perp^\be\ll\s_\perp,\qquad \frac{k^{\bar{\ga}}_\perp}{|k^{\bar{\ga}}_3|}p_\perp^\be\approx \frac{p^0_\perp}{|k^0_3|}p_\perp^\be \ll\s_\parallel.
\end{equation}
Under these estimates $k_\perp^{\bar{\ga}}\approx p_\perp^0 \gg p_\perp^\be$ and one can replace $k_\perp^{\bar{\ga}}\rightarrow p_\perp^0$ in all the functions entering into the integrand of \eqref{A_ampl1} except for $I_{lm}(\vf^\be_p)$.

The first inequality in \eqref{de_ga_region} is satisfied for any values of $\de_{\bar{\ga}}$ provided the estimates \eqref{est_for_pure} hold. Let us consider in detail the second inequality in \eqref{de_ga_region}, where
\begin{equation}
    E_{\al_0\be}=T_k^0+U_0-T^\be_k=\frac{(p_\perp^0)^2+(p_3^0)^2}{2 m_0} +U_0 - \frac{(p_\perp^\be)^2+(\bar{p}_3^\be)^2}{2 m_*}.
\end{equation}
Inasmuch as the conduction electrons in the crystal are distributed with respect to momenta in accordance with the Fermi-Dirac distribution, $p_\perp^\be\approx|\bar{p}_3^\be|\approx p_\be$, $p_\be$ is found from \eqref{electron_mom_in_cr}, and $T_k^\be$ is of order of the temperature for semiconductors and of order of the Fermi energy for conductors. We see that the contribution to the second inequality in \eqref{de_ga_region} of the kinetic energy of electron in the crystal can be discarded in comparison with $p_\perp^0$ and $p_\perp^\beta$. Then this inequality is fulfilled for any $\de_{\bar{\ga}}$ when
\begin{equation}
    p^0_\perp+p^\be_\perp\leqslant T_k^0+U_0.
\end{equation}
Neglecting $p^\be_\perp$ as compared to $p^0_\perp$, we have
\begin{equation}\label{p30_est}
    p^0_\perp\lesssim T_k^0+U_0.
\end{equation}
This inequality can always be satisfied for sufficiently large $p_3^0$. Thus, the second inequality in \eqref{de_ga_region} is satisfied with good accuracy for any values of $\de_{\bar{\ga}}$ so long as estimates \eqref{est_for_pure} hold true and $p_3^0$ obeys inequality \eqref{p30_est}.

Let us now consider the behavior of $I_{lm}(\vf^\be_p)$ at small $p_\perp^\be$. Substituting \eqref{change_var} into \eqref{Ilm0} and developing the result as a Taylor series with respect to $p_\perp^\be$, we have
\begin{equation}
\begin{split}
    p_\perp^\be \sin\de_{\bar{\ga}} I_{lm}(0)=&\,\frac{2}{p_\perp^0}\Big\{\cos((l-m)\de_{\bar{\ga}})+\\
    &\,+\big[\cos\de_{\bar{\ga}} \cos((l-m)\de_{\bar{\ga}}) +(l+m)\sin\de_{\bar{\ga}} \sin((l-m)\de_{\bar{\ga}})  \big] \frac{p_\perp^\be}{2 p_\perp^0} +O\Big(\frac{(p_\perp^\be)^2}{(p_\perp^0)^2}\Big)\Big\},
\end{split}
\end{equation}
where the integration measure has been taken into account. Taking $k_\perp^{\bar{\ga}}$ in the form \eqref{change_var}, substituting it into the integrand of \eqref{A_ampl2}, and expanding with respect to $p_\perp^\be\cos\de_{\bar{\ga}}$, we obtain for the remaining functions in the integrand of \eqref{A_ampl2}:
\begin{equation}
    a^\s_0+a^\s_1 p_\perp^\be \cos\de_{\bar{\ga}} +O\big((p_\perp^\be)^2\big),
\end{equation}
where $a_0$ and $a_1$ do not depend on $\de_{\bar{\ga}}$. Integrating over $\de_{\bar{\ga}}\in[0,\pi]$, we arrive at
\begin{equation}
    \mathcal{A}_{\al_0\be}=2\pi \theta(E_{\al_0\be}) \frac{E_{\al_0\be}}{p^0_\perp} e^{i(l_0-m)\vf^\be_p} \sum_{\s=-1}^1 \Big[a_0^\s \de_{l_0m} +\big((l_0-\s)a_0^\s +(l_0-m) p_\perp^\al a_1^\s \big) (\de_{l_0,m+1} -\de_{l_0,m-1})\frac{p_\perp^\be}{2p_\perp^0}+\cdots \Big].
\end{equation}
We see that the selection rule $l_0=m$ is violated by the small contributions of relative order $p_\perp^\be/p^0_\perp$. Contributions coming with the higher powers of $p_\perp^\be/p^0_\perp$ contain the terms with larger deviation from the equality $l_0=m$, the contributions of order $(p_\perp^\be/p^0_\perp)^k$ being such that $|l_0-m|\leqslant k$. Notice also that as thermionic emission is assumed to be negligibly small, one can suppose that $E_\be<0$ and, consequently, $E_{\al_0\be}>0$. This allows one to set $\theta(E_{\al_0\be})=1$ in the amplitude $\mathcal{A}_{\al_0\be}$ and to discard the second term in the square brackets in the differential probability \eqref{probab_diff}.

\begin{table}[t]
	\centering
	\begin{tabular}{cM{0.42\linewidth}M{0.42\linewidth}}
		\toprule
		 & $k_{3} = -5  $ eV &	$ k_{3} = -15 $ eV \\
		\midrule
		$ m = 1$  & \includegraphics[width=\linewidth]{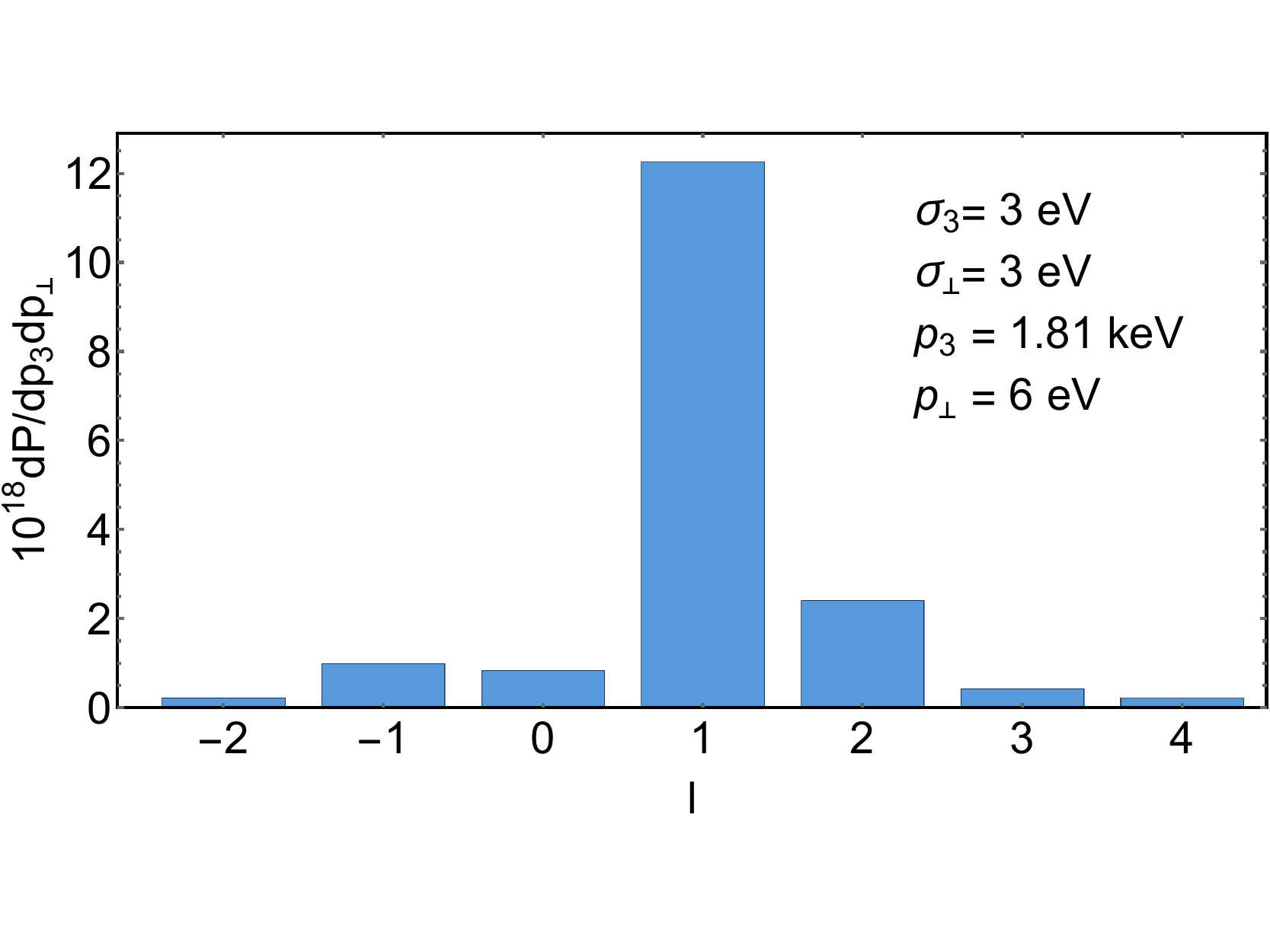} & \includegraphics[width=\linewidth]{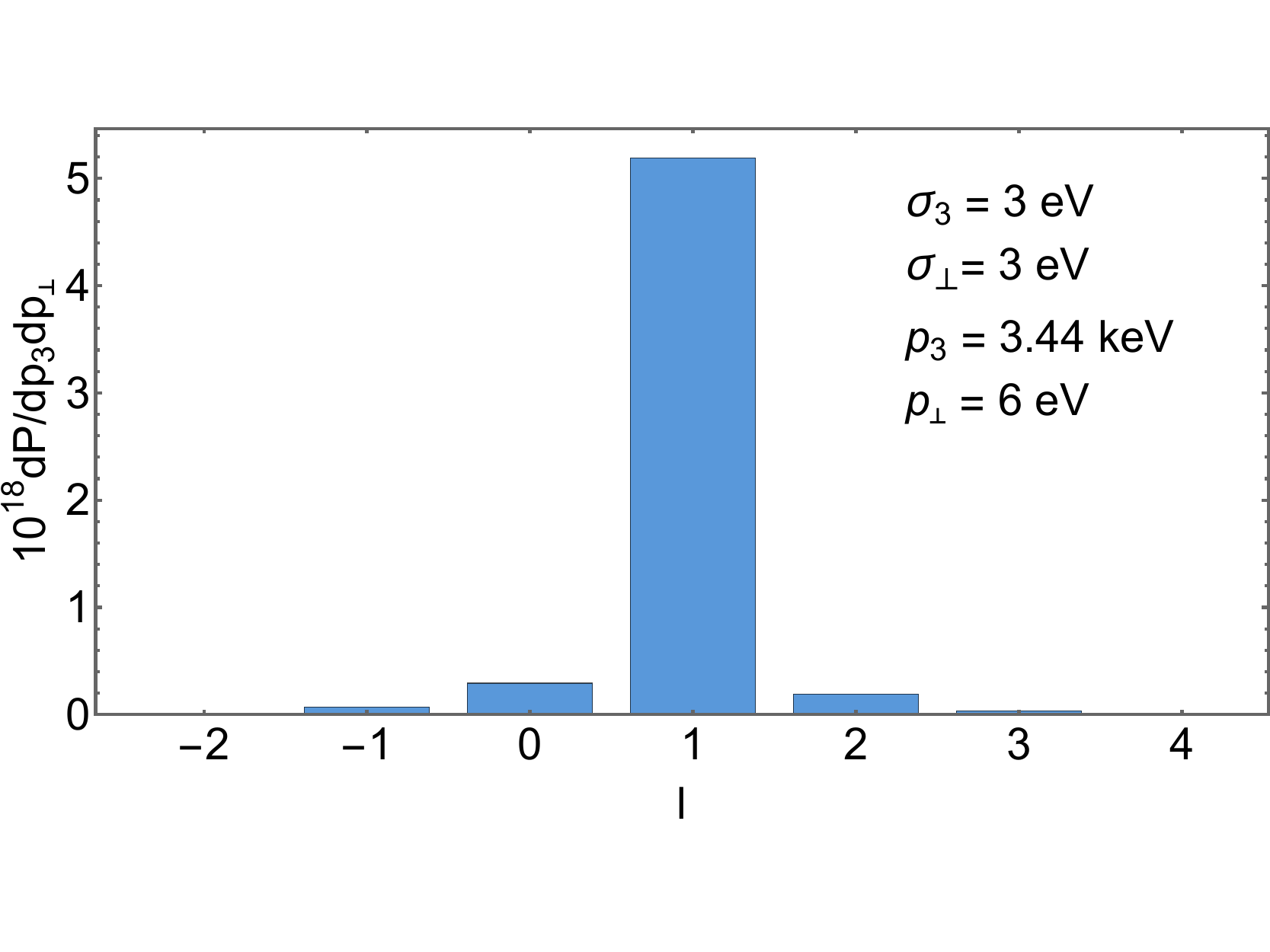} \\
		$ m = 16$ & \includegraphics[width=\linewidth]{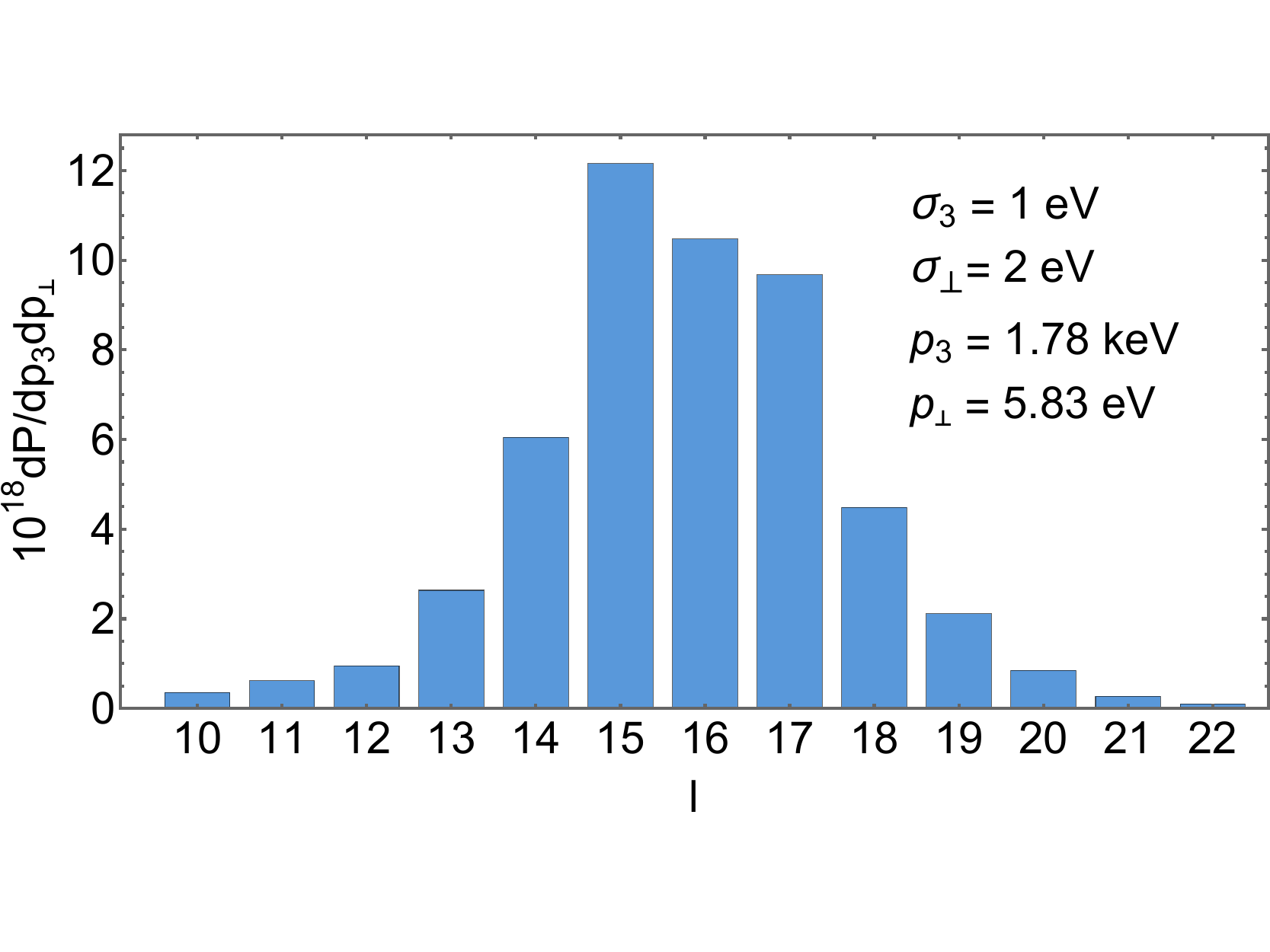} & \includegraphics[width=\linewidth]{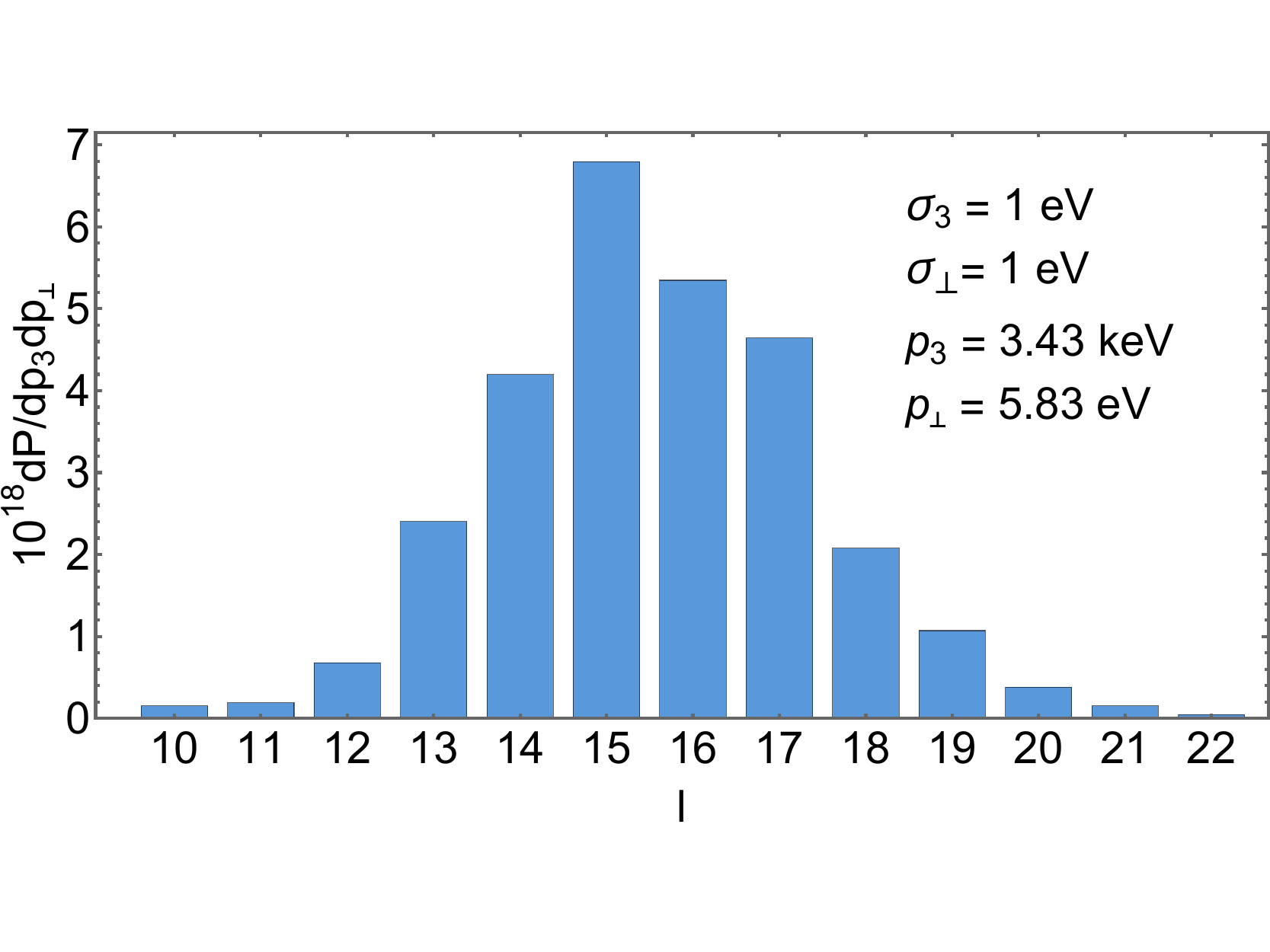} \\
		\bottomrule
	\end{tabular}
	\caption{{\footnotesize The differential probability \eqref{probab_diff} to record a twisted photoelectron with the projection of  angular momentum $l$ per photon for n-InSb at a temperature of $T = 2.5$ K. The state of the incident twisted photon is described by the Laguerre-Gaussian mode \eqref{GL_modes}. The dielectric permittivity of InSb used in calculations is presented in Fig. \ref{EPS_plot}. The parameters of the surface layer: $\de_+=0.02$ {\AA} and $\de_-=1.98$ {\AA} and so $\de_++\de_-=2$ {\AA}.}}
	\label{Table_2.5K}
\end{table}

Keeping only the leading contribution to the differential probability to record a twisted photoelectron \eqref{probab_diff}, we obtain
\begin{equation}\label{probab_diff1}
    \frac{dP(\al_0)}{dp_\perp^0 dp_3^0}=\frac{2\al}{p_\perp^0} \de_{l_0m} \int_0^\infty d p^\be_\perp d\bar{p}^\be_3 p^\be_\perp  \frac{\theta(-E_\be) E_{\al_0\be}}{e^{\be_T(E_\be-\mu)}+1} \Big|\sum_{\s=-1}^1 a_0^\s\Big|^2,
\end{equation}
where $E_\be=T_k^\be-U_0$. It follows from formula \eqref{probab_diff1} that all the photoelectrons have the orbital angular momentum projection $l_0=m$, in the approximation we consider. Moreover, inequality \eqref{p30_est} implies that the twisted photoelectrons are paraxial: Their perpendicular momentum is less than several eVs, whereas the longitudinal momentum component, $p_3\approx (2m_0 T^0_k)^{1/2}$, is of order of several keVs.

Now we discuss when the estimate,
\begin{equation}\label{main_est}
    p_\perp^0\approx k_\perp^{\bar{\ga}}\gg p_\perp^\be,
\end{equation}
can be satisfied. Given the fact that $k_\perp^{\bar{\ga}}\lesssim k_0^{\bar{\ga}}$, and $k_0^{\bar{\ga}}$ should not be much larger than the electronic work function, which is a quantity of order $1-10$ eV, the estimate \eqref{main_est} cannot be valid for metals. For example, for Cu at the Fermi energy $p^\be_\perp\approx p^\be=2.7$ keV. Nevertheless, the estimate \eqref{main_est} can be justified for semiconductors at small temperatures with conduction electrons possessing a small effective mass. Notice that for materials with anisotropic electron dispersion law only the components of the effective mass tensor perpendicular to the vector normal to the crystal interface are important for the estimates \eqref{main_est} to hold. In semiconductors, the kinetic energy of conduction electrons is of the same order of magnitude as the temperature.

%see fig.

%%%%%%%%figure here
\begin{figure}[tp]
\centering
\includegraphics*[width=0.45\linewidth]{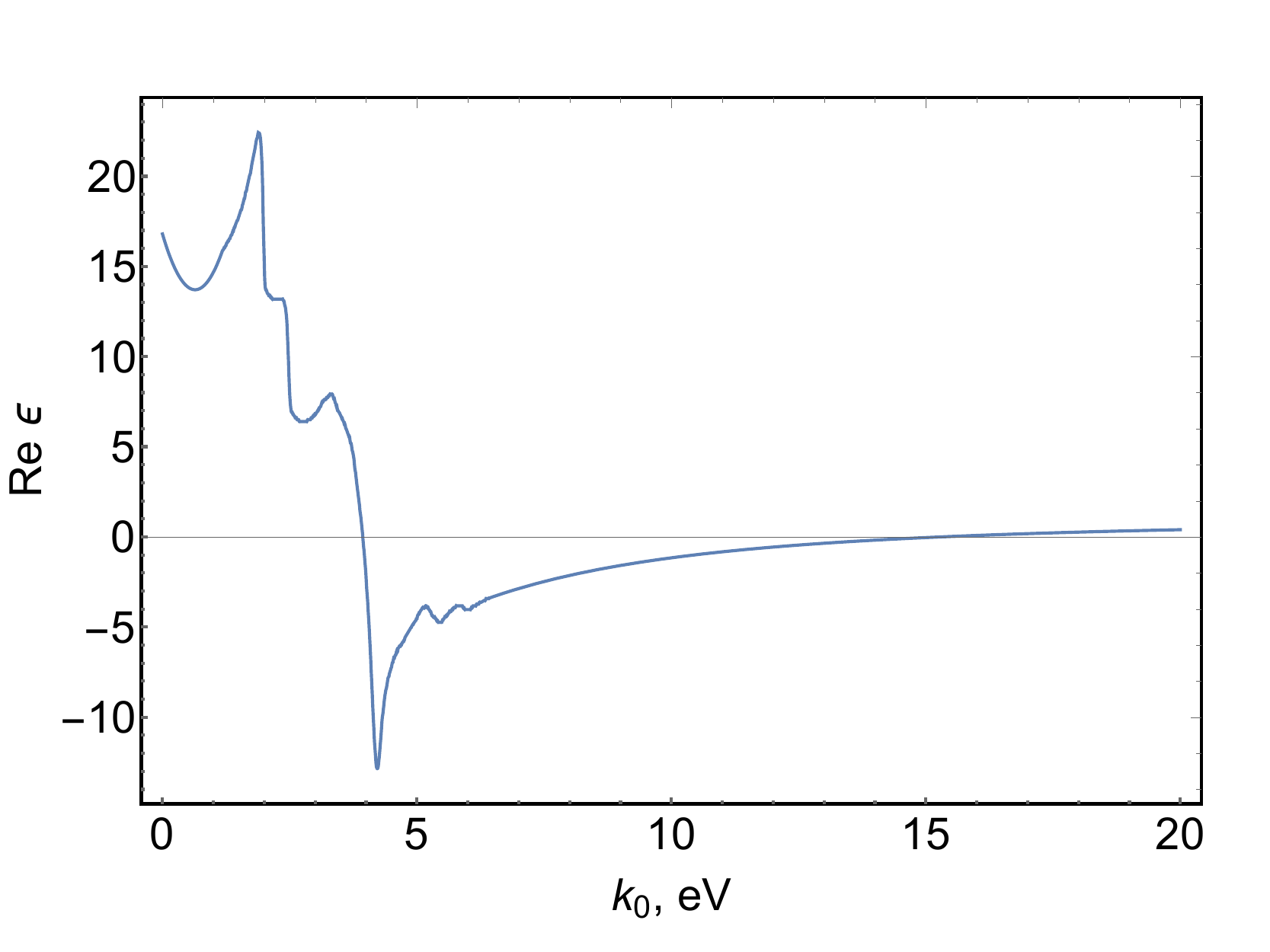}\;
\includegraphics*[width=0.45\linewidth]{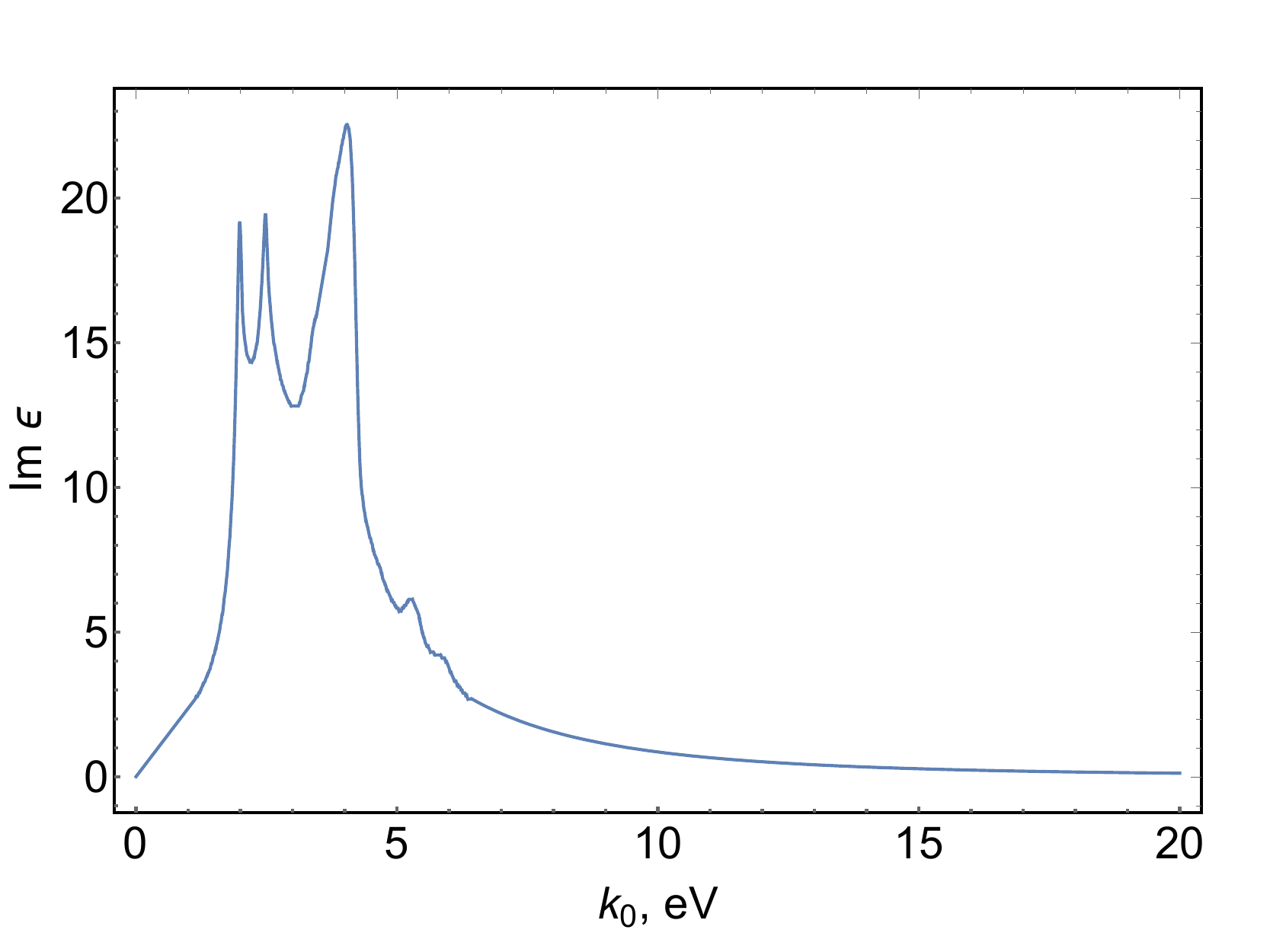}
\caption{{\footnotesize The real and imaginary parts of the dielectric permittivity $\e(k_0)$ of InSb used in calculations. The dielectric permittivity of InSb is taken from \cite{Kim2013} and is extrapolated to zero photon energy to coincide with static dielectric constant $\e_0$ and to higher photon energies by the Drude formula with the plasma frequency $\omega_p=15.8$ eV and the damping parameter $\ga=3.95$ eV.}}
\label{EPS_plot}
\end{figure}
%%%%%%%%figure here

The most optimal semiconductor for the estimate \eqref{main_est} to be valid is InSb which possesses a very small effective electron mass in the conduction band. At the temperature of evaporating liquid helium, $2.5$ K corresponding to $0.215$ meV, the modulus of momenta of conduction electrons in InSb is of order $1.76$ eV. The photon energies are of the order of or greater than the electronic work function $U_0=4.59$ eV. Of course, at such temperatures the concentration of conduction electrons in pure InSb is very small due to exponential suppression. In the approximation of nondegenerate electron gas, it is equal to
\begin{equation}
    n_e\approx Q_e e^{-\be_T E_g/2},\qquad Q_e:=2\Big(\frac{m_*}{2\pi\be_T}\Big)^{3/2}=2.51\times10^{19}\Big(\frac{m_*}{m_0} \frac{\be^{-1}_T[\text{K}]}{300} \Big)^{3/2}\;[\text{cm}{}^{-3}],
\end{equation}
where $E_g$ is the energy gap between the valence and conduction bands. At zero temperature, $E_g=0.24$ eV. However, if InSb contains shallow donor impurities of Se, S, or Te, then $E_g$ in the exponent can be approximately replaced by the ionization energy, $\e_i$, of such impurities, which is of order $0.7$ meV \cite{Ioffe_Inst}, and
\begin{equation}
    n_e\approx\sqrt{n_d Q_e}e^{-\be_T\e_i/2},
\end{equation}
where $n_d$ is the concentration of donor impurities. As a result, for a temperature of $2.5$ K, there is no exponential suppression of the number of electrons in the conduction band. A more accurate estimate is obtained if we abandon the approximation of nondegenerate electron gas. Then the electroneutrality condition has the form
\begin{equation}
    -Q_e \Li_{3/2}(-e^{\be_T \tilde{\mu}}) =\frac{n_d}{1+e^{-\be_T (-\e_i-\tilde{\mu})}}=n_e,
\end{equation}
where $\tilde{\mu}=\mu+U_0$ is the chemical potential with zero level at the bottom of the conduction band and $\Li_\nu(z)$ is the polylogarithm function. For a lightly doped crystal of InSb with $n_d=10^{14}$ cm${}^{-3}$ and the above parameters, we obtain $\tilde{\mu}=-0.225$ meV, $n_e=9.95\times10^{12}$ cm${}^{-3}$, and $1/r_D=1.39$ eV. The Debye radius has been found with the aid of the standard formula \eqref{Debye_radius}. For such electron concentrations, the parameter characterizing the overlap of electron wave functions bound to impurities $n_e^{1/3} \e_0/(\al m_{eff})=0.14$ implying that the semiconductor is lightly doped and the impurity band is not formed \cite{ShklovEfr_book1984}. The ratio of energy of the screened Coulomb interaction to kinetic energy of electrons can be estimated as
\begin{equation}\label{strenght_CI}
	\al n_e^{1/3} e^{-n_e^{-1/3}/r_D} \be_T/\e_0=0.032.
\end{equation}
Hence the approximation of free conduction electrons is justified. The probability to detect a twisted photoelectron in this case is shown in the figures in table \ref{Table_2.5K}. By decreasing the temperature, one can achieve the fulfillment of estimates \eqref{est_for_pure} for lower photon energies. For example, at a temperature of $0.5$ K, the modulus of momenta of conduction electrons in InSb is of order $0.79$ eV. For impurity concentration $n_d=5\times 10^{14}$ cm${}^{-3}$, we have $\tilde{\mu}=-0.238$ meV, $n_e=1.11\times10^{10}$ cm${}^{-3}$, and $1/r_D=0.104$ eV. The parameter $n_e^{1/3} \e_0/(\al m_{eff})=0.014$ and so the semiconductor is lightly doped. The parameter \eqref{strenght_CI} is $0.042$.

%The probability to detect a twisted photoelectron in this case is presented in Figs. \ref{Table_0.5K}.

%\begin{table}
%	\centering
%	\begin{tabular}{cM{0.42\linewidth}M{0.42\linewidth}}
%		\toprule
%		& $k_{30} = -5  $ eV &	$ k_{30} = -15 $ eV \\
%		\midrule
%		$ m = 1$  & \includegraphics[width=\linewidth]{2_1.pdf} & \includegraphics[width=\linewidth]{2_3.pdf} \\
%		$ m = 16$ & \includegraphics[width=\linewidth]{2_2.pdf} & \includegraphics[width=\linewidth]{2_4.pdf} \\
%		\bottomrule
%	\end{tabular}
%	\caption{{\footnotesize The same as in Table \ref{Table_2.5K} but for the temperature $T = 0.5$ K.}}
%	\label{Table_0.5K}
%\end{table}

The temperature at which the estimate \eqref{main_est} is valid can be increased if instead of the ordinary semiconductors one uses the bulk Dirac or Weyl semimetals where the electrons have a linear dispersion law $T_k=v_F |\spp|$ \cite{Dimmock1966,Yusheng1985III,Wehling2014,Ali2014,Crassee2018,Lee2015,Jenkins2016,Bernardo2020,Regmi2024}. Of course, for such materials the formalism developed above for the description of photoelectric effect changes a little but these changes do not affect the conditions \eqref{est_for_pure} that validate a complete transfer of the angular momentum projection of the photon to the photoelectron. Taking as an estimate the value of $v_F$ for graphene, $v_F\approx1/300$, we see that already at the temperature of evaporating liquid nitrogen $63.5$ K the modulus of momentum of conduction electrons turns out to be equal to $1.64$ eV that should be compared with the electronic work function. For pure graphene the electronic work function $U_0=4.5$ eV. The concentration of electrons in the conduction band reads
\begin{equation}
    n_e=\frac{12\pi \zeta(3)}{(2\pi\be_T v_F)^3}\approx4.11\times 10^{8}\Big(\frac{c}{v_F} \frac{\be^{-1}_T[\text{K}]}{300}\Big)^3\; [\text{cm}{}^{-3}].
\end{equation}
For the above parameters, $n_e\approx1.05\times10^{14}$ cm${}^{-3}$. Apparently, these estimates will not change significantly for two-dimensional materials with a linear electron dispersion law. However, the description of photoelectric effect caused by twisted photons in such materials requires a more serious modification of the model and will be carried out in a separate paper.

%отметить, что кристалл должен быть достаточно чистым (оценка для примесей) с ровной поверхностью (это важно, для сохранения инвариантности поверхности относительно поворотов вокруг нормали)

% отметить, что химический потенциал можно уменьшать за счет добавления компенсирующих примесей
% мы рассматриваем слабо легированный InSb, чтобы пренебречь взаимодействие электронов с примесями и между собой

%эффективная масса электрона в зоне проводимости в InSb: 0.014m_0
%сродство электрона в InSb: 4.59 эВ
%статическая диэлектрическая проницаемость в InSb: 16.8
%статическая диэлектрическая проницаемость в Cd3As2: 36
%сродство электрона в чистом графене: 4.5 эВ
%скорость Ферми в чистом графене и Cd3As2: 1/300
%энергия ферми в Cu: 7.0 эВ
%работа выхода в Cu: 4.36 эВ. Следовательно, U_0=7.0+4.36=11.36 эВ
%эффективная масса электрона в зоне проводимости в Cu: m_0
%важны компоненты тензора масс, перпендикулярные к вектору нормали к поверхности.

%ширина запрещенной зоны в InSb: 0.24 - 6 10^(-4) tt^2/(tt + 500), tt -- температура в кельвинах
%при малых температурах mu\approx(E_c+E_v)/2
%энергия ионизации неглубоких донорных примесей Se,S,Te порядка 7x10^{-4} эВ
%при малых температурах mu\approx(E_c+E_d)/2

%на рисунках приведена вероятность на один падающий фотон

\section{Conclusion}

Let us sum up the results. We have developed the theory of the surface photoelectric effect induced by twisted photons in the effective mass approximation. We have obtained the explicit expression \eqref{probab_diff} for the differential probability to detect an electron with definite projection of the orbital angular momentum onto the normal to crystal surface, definite momentum projection onto this normal, and definite modulus of the perpendicular momentum component. We have found the restrictions on the momenta of electrons in the crystal \eqref{est_for_pure}, \eqref{p30_est} whose fulfillment is necessary for the ejected photoelectrons to be twisted. These restrictions can be satisfied for lightly doped semiconductors and Dirac or Weyl semimetals at sufficiently low temperatures and small effective masses of conduction electrons.

%, \ref{Table_0.5K}

We have shown that the crystal of InSb lightly doped by shallow donors at temperatures lower than $2.5$ K with interface without defects can be used as the pure source of twisted photoelectrons when irradiated by twisted photons. The distributions of ejected photoelectrons with respect to the projection of orbital angular momentum are presented in the figures in table \ref{Table_2.5K}. The estimates \eqref{est_for_pure}, \eqref{p30_est} are also fulfilled at temperatures lower than $60$ K for the Dirac and Weyl semimetals with Fermi velocity of order $c/300$ and the chemical potential near the top of the Dirac cone. This suggests that these materials can also be employed for production of twisted electrons by the surface photoelectric effect.

In the present paper, we have not considered the photoelectric effect in two-dimensional materials. This is the aim of our future research. However, we expect that the restrictions \eqref{est_for_pure}, \eqref{p30_est} will remain the same and these materials can also be used as a source of twisted electrons when the estimates \eqref{est_for_pure}, \eqref{p30_est} are satisfied.

%\newpage
\appendix
\section{Mode functions of an electron quantum field}\label{Electron_Modes_App}

Let us now find the mode functions of the electrons. As discussed in Sec. \ref{Hamiltonian_sec}, the electron mode functions satisfy Eq. \eqref{Schroed_eqn} with the matching conditions for the envelope function (see, e.g., \cite{Rodina2002} for details)
\begin{equation}\label{join_conds_e}
    \psi_\al\big|_{z=-\de_-/2}= \psi_\al\big|_{z=\de_+/2},\qquad \frac{1}{m_*}\partial_z\psi_\al\big|_{z=-\de_-/2}= \frac{1}{m_0} \partial_z\psi_\al\big|_{z=\de_+/2}.
\end{equation}
These boundary conditions ensure the continuity of the electron current density averaged over the unit cell at the crystal boundary. Since Eq. \eqref{Schroed_eqn} is invariant under translations in the $(x,y)$ plane, its solution has the form
\begin{equation}\label{electron_mod0}
    \psi_\al(\spx)=e^{i\spp_\perp\spx_\perp}\psi_\al(z),
\end{equation}
and
\begin{equation}\label{Schroed_eqn_1}
    \big[-\frac{\partial}{\partial z} \frac{1}{2\tilde{m}} \frac{\partial}{\partial z}+U(z) \big]\psi_\al(z)=\big(E-\frac{p_\perp^2}{2\tilde{m}}\big)\psi_\al(z),
\end{equation}
where the effective electron mass depends on $z$ as discussed in Sec. \ref{Hamiltonian_sec}. The normalization condition for the functions \eqref{electron_mod0} is written as
\begin{equation}
    \int_{-L/2}^{L/2} dz|\psi_\al(z)|^2=L,
\end{equation}
where $L$ is a sufficiently large dimension of the box where the electron is confined.

On substituting the explicit expression for the potential \eqref{Uz} with $U_{ext}(z)=0$ into the Schr\"{o}dinger equation \eqref{Schroed_eqn_1}, we obtain a confluent hypergeometric equation for $z>\de_+/2$. For $z<-\de_-/2$, the Schr\"{o}dinger equation is not exactly solvable. When the Debye radius, $r_D$, and the electron momentum in the crystal are sufficiently large,  $|\bar{p}_3| r_D\gg1$, one can neglect the Debye screening and replace the Yukawa potential by the Coulomb potential. Recall that in the approximation of nondegenerate electron gas,
\begin{equation}\label{Debye_radius}
    r_D=\sqrt{\frac{\e_0}{4\pi\al\be_Tn_e}},
\end{equation}
where $\al$ is the fine structure constant and it is assumed that the free charge carriers are the electrons in the conduction band. The Debye radius increases with decreasing temperature in semiconductors due to a rapid decrease of the concentration of electrons in the conduction band. Below we derive the electron mode functions in the case when $r_D\rightarrow\infty$ and at the end of this appendix we discuss how to account for the finiteness of the Debye radius.

In the limit $r_D\rightarrow\infty$, it is convenient to choose a complete set of solutions as
\begin{equation}\label{electron_mod1}
    \psi_\al(z)=C_\al
    \left\{
      \begin{array}{ll}
        W_{\frac{i\vk}{2\eta p_3},\frac12}(-2i\eta p_3z), & \hbox{with $z>\de_+/2$;} \\[1em]
        A_\al W_{-\frac{i\bar{\vk}}{2 \bar{p}_3},\frac12}(2i \bar{p}_3z) +B_\al W_{\frac{i\bar{\vk}}{2 \bar{p}_3},\frac12}(-2i \bar{p}_3z), & \hbox{with $z<-\de_-/2$,}
      \end{array}
    \right.
\end{equation}
where $A_\al$, $B_\al$, and $C_\al$ are some constants, $\eta=\pm1$ characterizes the direction of wave propagation, $W_{a,b}(z)$ is the Whittaker function,
\begin{equation}\label{vk_bvk_p3_bp3}
\begin{aligned}
    \vk&=\frac{\al m_0}{2}\frac{\e_0-1}{\e_0+1},&\qquad \bar{\vk}&=\frac{\al m_*}{2\e_0}\frac{\e_0-1}{\e_0+1}, \\
    p_3&=\sqrt{2m_0E_\al-p_\perp^2},&\qquad\bar{p}_3&=\sqrt{2m_*(E_\al+U_0) -p_\perp^2}.
\end{aligned}
\end{equation}
In these expressions, the branch of the root with nonnegative imaginary part is chosen. For bound states in the crystal, the solution has the form \eqref{electron_mod1} with $\eta=1$. The quantities $2\vk$ and $2\bar{\vk}$ are the Bohr momenta of the electron in vacuum and in the crystal, and $p_3$ and $\bar{p}_3$ specify the electron momenta at $z\rightarrow\infty$ and $z\rightarrow-\infty$, respectively.

For $z\rightarrow+\infty$, there is an asymptotics \cite{GrRy}
\begin{equation}
    W_{\frac{i\vk}{2\eta p_3},\frac12}(-2i\eta p_3z)\simeq e^{i\eta p_3z} (-2i\eta p_3 z)^{i\vk/(2\eta p_3)}(1+O(1/z)).
\end{equation}
The leading term of the asymptotics accurately approximates the exact expression when
\begin{equation}
    |z|\gg \max(\vk^2/|p_3|^3, \vk/|p_3|^2).
\end{equation}
In the case $p_3>0$, the value $\eta=1$ corresponds to a wave running to the right, and $\eta=-1$ refers to a wave running to the left. The $z$-dependent factor in front of the exponent is the so-called Coulomb logarithm. For the energies $E_\al<0$, there are bound states of the electron in the plate. For these states $\eta=1$ and so, bearing in mind the chosen branch of the root in \eqref{vk_bvk_p3_bp3}, they decay exponentially in the limit $z\rightarrow+\infty$.

In order to impose the matching conditions \eqref{join_conds_e}, it is convenient to use the expansion of the Whittaker function at $z\rightarrow0$,
\begin{equation}\label{Whittaker_z0}
    W_{\frac{i\vk}{2\eta p_3},\frac12}(-2i\eta p_3z)= \frac{1}{\Ga\big(1-\frac{i\vk}{2\eta p_3}\big)}\Big\{1 -\big[\psi\big(1-\frac{i\vk}{2\eta p_3}\big) +\ln\big(-2i\eta p_3 z e^{2\ga_E -1}\big) -\frac{i\eta p_3}{\vk}\big] \vk z\Big\} +\cdots,
\end{equation}
where $\psi(z)$ is a digamma function and the principal branches of multivalued functions are taken. The expression \eqref{Whittaker_z0} accurately approximates the exact solution provided $|z|\ll\min(1/|p_3|,1/\vk)$. Supposing that this estimate is valid for $|z|=\de_\pm/2$, the conditions \eqref{join_conds_e} imply
\begin{equation}\label{A_B_coeff}
\begin{split}
    A_\al&=-\frac{\Ga\big(1+\frac{i\bar{\vk}}{2\bar{p}_3}\big)}{\Ga\big(1-\frac{i\vk}{2\eta p_3}\big)} \frac{ \psi\big(1-\frac{i\bar{\vk}}{2\bar{p}_3}\big) +\ln\big(i\bar{p}_3\de_- e^{2\ga_E}\big) -\frac{i\bar{p}_3}{\bar{\vk}} -\frac{\vk}{\la\bar{\vk}}\big[ \psi\big(1-\frac{i\vk}{2\eta p_3}\big) +\ln\big(-i\eta p_3\de_+ e^{2\ga_E}\big) -\frac{i\eta p_3}{\vk} \big]}{i\pi\cth\frac{\pi\bar{\vk}}{2\bar{p}_3}+\ln(-i\bar{p}_3) -\ln(i\bar{p}_3)},\\
    B_\al&=-\frac{\Ga\big(1-\frac{i\bar{\vk}}{2\bar{p}_3}\big)}{\Ga\big(1-\frac{i\vk}{2\eta p_3}\big)} \frac{ \psi\big(1+\frac{i\bar{\vk}}{2\bar{p}_3}\big) +\ln\big(-i\bar{p}_3\de_- e^{2\ga_E}\big) +\frac{i\bar{p}_3}{\bar{\vk}} -\frac{\vk}{\la\bar{\vk}}\big[ \psi\big(1-\frac{i\vk}{2\eta p_3}\big) +\ln\big(-i\eta p_3\de_+ e^{2\ga_E}\big) -\frac{i\eta p_3}{\vk} \big]}{-i\pi\cth\frac{\pi\bar{\vk}}{2\bar{p}_3}+\ln(i\bar{p}_3) -\ln(-i\bar{p}_3)},
\end{split}
\end{equation}
where $\la:=m_0/m_*$. The following relation is valid
\begin{equation}\label{A_B_rel}
    A_\al\big|_{\bar{p}_3\rightarrow-\bar{p}_3}=B_\al.
\end{equation}
Due to the fact that we are considering the case when $\bar{p}_3>0$, the expression in the denominator of \eqref{A_B_coeff} can be simplified:
\begin{equation}
    \ln(-i\bar{p}_3) -\ln(i\bar{p}_3)=-i\pi.
\end{equation}
However, after such a substitution in \eqref{A_B_coeff}, the symmetry relation \eqref{A_B_rel} does not hold. Notice also that $\vk/(\la \bar{\vk})=\e_0$. For a metal plate, one needs to take the limit $\bar{\vk}\rightarrow0$. In this case,
\begin{equation}
\begin{split}
    A_\al&= -\frac{i\vk}{2\bar{p}_3} \frac{\psi\big(1-\frac{i\vk}{2\eta p_3}\big) +\ln\big(-i\eta p_3\de_+ e^{2\ga_E}\big) -i\frac{\eta p_3-\la\bar{p}_3}{\vk}}{\la\Ga\big(1-\frac{i\vk}{2\eta p_3}\big)},\\
    B_\al&=\frac{i\vk}{2\bar{p}_3} \frac{\psi\big(1-\frac{i\vk}{2\eta p_3}\big) +\ln\big(-i\eta p_3\de_+ e^{2\ga_E}\big) -i\frac{\eta p_3+\la\bar{p}_3}{\vk}}{\la\Ga\big(1-\frac{i\vk}{2\eta p_3}\big)}.
\end{split}
\end{equation}
The normalization condition for the scattering states can be cast into the form
\begin{equation}\label{norm_cond1}
    \frac{C^2_\al}{2} \big[e^{\frac{\pi\vk}{2 p_3}} +e^{-\frac{\pi\bar{\vk}}{2 \bar{p}_3}} (|A_\al|^2+|B_\al|^2) \big]=1.
\end{equation}
The normalization condition for the states localized in the plate reduces to
\begin{equation}\label{norm_cond2}
    \frac{C^2_\al}{2} e^{-\frac{\pi\bar{\vk}}{2 \bar{p}_3}} (|A_\al|^2+|B_\al|^2) =1.
\end{equation}
In deriving relations \eqref{norm_cond1} and \eqref{norm_cond2}, it is assumed that
\begin{equation}
\begin{gathered}
    L\gg \frac{1}{p_3},\qquad L\gg\frac{\vk^2}{p_3^3},\\
    L\gg \frac{1}{\bar{p}_3},\qquad L\gg \frac{\bar{\vk}^2}{\bar{p}_3^3},\qquad L\gg \frac{\bar{\vk}^2}{\bar{p}_3^3} e^{\frac{\pi\bar{\vk}}{2\bar{p}_3}}.
\end{gathered}
\end{equation}
For large $\pi\bar{\vk}/(2\bar{p}_3)$, the last condition cannot be met for $L$ reasonable from the physical point of view. For such $\bar{p}_3$, it is necessary to take into account the Debye screening, finiteness of the crystal, defects in it, and interaction leading to decoherence of the electron wave function.

The states \eqref{electron_mod0} form a complete set. In order to find the spectral measure entering into the completeness relation, one can assume that the states localized in the crystal, i.e., with $E_\al<0$, vanish at $z=-L/2$, whereas the scattering states, i.e., the states with $E_\al>0$, satisfy the periodic boundary conditions at $z=\pm L/2$. Then, making use of the argument principle in the complex energy plane and finding the leading asymptotics of the spectral density of states for $L\rightarrow\infty$, we derive
\begin{equation}
    \sum_\al=\int\frac{V d\spp^\al_\perp}{(2\pi)^3} \Big[ \int_{-U_0}^0 \frac{dE_\al \theta\big(E_\al+U_0-(p_\perp^\al)^2/(2m_*)\big) m_* }{\sqrt{2m_*(E_\al+U_0)-(p_\perp^\al)^2}} +\sum_{\eta=\pm1}\int_0^\infty \frac{dE_\al \theta\big(E_\al-(p_\perp^\al)^2/(2m_0)\big) m_0 }{\sqrt{2m_0E_\al-(p_\perp^\al)^2}} \Big].
\end{equation}
It is convenient to rewrite these integrals in terms of momenta
\begin{equation}\label{sum_al}
    \sum_\al=\int\frac{V d\spp^\al_\perp}{(2\pi)^3} \Big[ \int_0^\infty  d\bar{p}^\al_3 \theta(-E_\al) +\int_{-\infty}^\infty dp^\al_3 \Big],
\end{equation}
where $E_\al=[(p^\al_\perp)^2+(\bar{p}^\al_3)^2]/(2 m_*)-U_0$ in the first term and $E_\al=[(p^\al_\perp)^2+(p^\al_3)^2]/(2 m_0)$ in the second term.

As long as the photon mode functions are linear combinations of exponents $\exp(\pm ik_3z)$ (see Appendix \ref{Photon_Modes_App}), the evaluation of the integral \eqref{wbg_baa}, \eqref{wbg_baa1} entering into the transition amplitude $\mathcal{A}_{\al_0\be}$ is reduced to the evaluation of the integrals
\begin{equation}\label{De_integrl}
    \De(a,b;p_1,p_2,k):=\int_0^\infty dz e^{-ikz} W_{a,\frac12}(2ip_1z) W_{b,\frac12}(2ip_2z).
\end{equation}
It seems that this integral cannot be expressed in term of known special functions. Nevertheless, it can be rewritten as an integral of a function that does not contain rapidly oscillating factors. To this end, one can use the integral representation of the product of two Whittaker functions (formula 9.225.2 of \cite{GrRy}). On substituting this representation into \eqref{De_integrl}, the integral over $z$ is performed and we obtain
\begin{equation}\label{De_integrl1}
    \De(a,b;p_1,p_2,k)=-\frac{4p_1p_2}{\Ga(1-a-b)} \int_0^\infty dt \frac{t^{-a-b} (t+2ip_1)^{a-1} (t+2ip_2)^{b-1}}{(t+ip_1+ip_2+ik)^2} F(1-a,1-b;1-a-b;\theta),
\end{equation}
where
\begin{equation}
    \theta:=\frac{t(t+2ip_1+2ip_2)}{(t+2ip_1)(t+2ip_2)},
\end{equation}
and $F(a,b;c;z)$ is the Gaussian hypergeometric function. The integral representation \eqref{De_integrl1} is useful for numerical calculations.

The integrals \eqref{wbg_baa1}, \eqref{wbg_baa1sigma} entering into the amplitude $\mathcal{A}_{\al_0\be}$ have the form
\begin{equation}
    w_{\s\be\al_0}^{\bar{\ga}}=p_\perp^{\be} \int_{-\infty}^\infty \frac{dz}{2\tilde{m}} \psi^*_{\be} \psi_{\al_0}\phi^*_{\bar{\ga},-\s},\qquad w_{0\be\al_0}^{\bar{\ga}}=-\int_{-\infty}^\infty dz\psi^*_\be\psi_{\al_0} \big[\frac{k_\perp^{\bar{\ga}}}{4\tilde{m}}(\phi^*_{\bar{\ga}-} +\phi^*_{\bar{\ga}+}) +ik_0^{\bar{\ga}}(\Phi^*_{\bar{\ga}})_3 \big],
\end{equation}
where $\s=\pm1$. Substituting the explicit expressions for the mode functions, we find
\begin{equation}\label{w_expl_s}
    w_{\s\be\al_0}^{\bar{\ga}}=p_\perp^{\be} C_\be C_{\al_0} a_{\bar{\ga}}\Big[\frac{\sum_{s'=\pm1} r_{s'}^* (f^{s'}_{-\s})^*}{2m_0} \De_+ +\frac{(\tilde{f}^s_{-\s})^*}{2m_0} \De_- +\frac{\sum_{s'=\pm1} l_{s'}^* (\tilde{g}^{s'}_{-\s})^*}{2 m_*}J \Big],
\end{equation}
where $\s=\pm1$ and, for brevity, the notation has been introduced
\begin{equation}\label{De_pm_J}
\begin{split}
    \De_\pm:=&\,\De\Big(-\frac{i\vk}{2(p_3^\be)^*},\frac{i\vk}{2p_3^0};(p_3^\be)^*,-p_3^0,\pm k_3^{\bar{\ga}}\Big),\\
    J:=&\,A^*_{\be} A_{\al_0} \De\Big(\frac{i\bar{\vk}}{2\bar{p}_3^\be},-\frac{i\bar{\vk}}{2\bar{p}_3^0};\bar{p}_3^\be,-\bar{p}_3^0, (\bar{k}_3^{\bar{\ga}})^*\Big) +A^*_{\be} B_{\al_0} \De\Big(\frac{i\bar{\vk}}{2\bar{p}_3^\be},\frac{i\bar{\vk}}{2\bar{p}_3^0};\bar{p}_3^\be,\bar{p}_3^0, (\bar{k}_3^{\bar{\ga}})^*\Big)+\\
    &+B^*_{\be} A_{\al_0} \De\Big(-\frac{i\bar{\vk}}{2\bar{p}_3^\be},-\frac{i\bar{\vk}}{2\bar{p}_3^0};-\bar{p}_3^\be,-\bar{p}_3^0, (\bar{k}_3^{\bar{\ga}})^*\Big) +B^*_{\be} B_{\al_0} \De\Big(-\frac{i\bar{\vk}}{2\bar{p}_3^\be},\frac{i\bar{\vk}}{2\bar{p}_3^0};-\bar{p}_3^\be,\bar{p}_3^0, (\bar{k}_3^{\bar{\ga}})^*\Big),
\end{split}
\end{equation}
and
\begin{equation}
    p^\be_3=\Big[2m_0\Big(\frac{(\bar{p}_3^\be)^2}{2m_*}+\frac{(p_\perp^\be)^2}{2m_*} -U_0\Big) -(p_\perp^\be)^2 \Big]^{1/2},\qquad \bar{p}^0_3=\Big[2m_*\Big(\frac{(p_3^0)^2}{2m_0}+\frac{(p_\perp^0)^2}{2m_0} +U_0\Big) -(p_\perp^0)^2 \Big]^{1/2}.
\end{equation}
Besides,
\begin{equation}\label{w_expl_0}
\begin{split}
    w_{0\be\al_0}^{\bar{\ga}} =&-\frac{k_\perp^{\bar{\ga}}}{2 p^\be_\perp}(w_{+\be\al_0}^{\bar{\ga}}+w_{-\be\al_0}^{\bar{\ga}}) -ik_0^{\bar{\ga}}  C_\be C_{\al_0} a_{\bar{\ga}}\Big[\frac{\sum_{s'=\pm1} r_{s'}^* (f^{s'}_{3})^*}{-ik_3^{\bar{\ga}}} (\De_+ -\De^0)+\\
    &+\frac{(\tilde{f}^s_{3})^*}{ik_3^{\bar{\ga}}} (\De_- -\De^0) +\frac{\sum_{s'=\pm1} l_{s'}^* (\tilde{g}^{s'}_{3})^*}{i(\bar{k}_3^{\bar{\ga}})^*}(J-J^0) \Big],
\end{split}
\end{equation}
where $\De^0$ and $J^0$ are obtained from \eqref{De_pm_J} by the replacement $k_3^{\bar{\ga}}=\bar{k}_3^{\bar{\ga}}=0$.

The states of the electron in the crystal that have been obtained above can be used when
\begin{equation}\label{Debye_estim}
    \bar{p}_3 r_D\gg1,\qquad z\geqslant -R,
\end{equation}
where $R:=7r_D$. For $z<-R$, one can suppose with good accuracy that the mode functions have the form
\begin{equation}\label{free_electr_wf}
    \tilde{A}_\al e^{-i\bar{p}_3z} +\tilde{B}_\al e^{i\bar{p}_3z},
\end{equation}
where
\begin{equation}
    \tilde{A}_\al=(-2i\bar{p}_3R)^{-i\bar{\vk}/(2\bar{p}_3)}A_\al,\qquad \tilde{B}_\al=(2i\bar{p}_3R)^{i\bar{\vk}/(2\bar{p}_3)} B_\al.
\end{equation}
If the first condition in \eqref{Debye_estim} is not met, then for such $\bar{p}_3$ the electron mode functions can be found numerically on the interval $z\in[-R,-\de_-/2]$, while outside of this interval they ought to be matched with the wave functions of the form \eqref{free_electr_wf}.

\section{Mode functions of a quantum electromagnetic field}\label{Photon_Modes_App}

Let us find the mode functions of the quantum electromagnetic field in the presence of a plate with dielectric permittivity $\e(k_0)$. For $\e(k_0)>1$, there are the mode functions that exponentially tend to zero when $z\rightarrow+\infty$. These are the bound states of photons in the plate corresponding to the total internal reflection. For the process we investigate -- the surface photoelectric effect -- these modes are not excited by the external electromagnetic waves in the leading order of perturbation theory. Hence, we do not take into account these modes in expansion \eqref{field_operators}. Then the modes of the electromagnetic field possessing nonzero limit for $z\rightarrow+\infty$ are written as
\begin{equation}\label{phot_modes}
    \bs\phi_{\ga}(\spx)=a_\ga e^{i\spk_\perp\spx_\perp}
    \left\{
      \begin{array}{ll}
        \sum_{s'=\pm1}r_{s'}\mathbf{f}_{s'} e^{i k_3z} +\tilde{\mathbf{f}}_s e^{-ik_3 z}, & \hbox{for $z>0$;} \\[1em]
        \sum_{s'=\pm1}l_{s'}\tilde{\mathbf{g}}_{s'} e^{-i \bar{k}_3z}, & \hbox{for $z<0$,}
      \end{array}
    \right.
\end{equation}
where
\begin{equation}\label{polarization_vects}
\begin{split}
    \mathbf{f}_{s'}&=(k_3\cos\vf_k-is'k_0\sin\vf_k,k_3\sin\vf_k+is'k_0\cos\vf_k,-k_\perp)/(\sqrt{2}k_0),\\
    \tilde{\mathbf{f}}_{s}&=(-k_3\cos\vf_k-isk_0\sin\vf_k,-k_3\sin\vf_k+isk_0\cos\vf_k,-k_\perp)/(\sqrt{2}k_0),\\
    \tilde{\mathbf{g}}_{s'}&=(-\frac{\bar{k}_3}{\e^{1/2}}\cos\vf_k-is'k_0\sin\vf_k,-\frac{\bar{k}_3}{\e^{1/2}}\sin\vf_k+is'k_0\cos\vf_k, -\frac{k_\perp}{\e^{1/2}})/(\sqrt{2}k_0).
\end{split}
\end{equation}
The indices $s=\pm1$, $s'=\pm1$ characterize the circular polarization, $\vf_k:=\arg(k_1+ik_2)$, $k_3:=\sqrt{k_0^2-k_\perp^2}$, $\bar{k}_3:=\sqrt{\e k_0^2-k_\perp^2}$, and such a branch of the square root is taken that has a nonnegative imaginary part. This choice of the square root ensures attenuation of the electromagnetic field modes for $z\rightarrow-\infty$ when $\bar{k}_3^2<0$ what is realized for $\e(k_0)<1$. The coefficients $r_{s'}$ and $l_{s'}$ are found from the matching conditions \eqref{joint_conds} and can be cast into the form
\begin{equation}
    r_{s'}=\frac{1}{2(\e-1)}\Big[\frac{(\e k_3-\bar{k}_3)^2}{\e k_3^2-k_\perp^2} -ss'\frac{(k_3-\bar{k}_3)^2}{k_0^2} \Big],\qquad l_{s'}=\frac{\e^{1/2}k_3}{\e k_3+\bar{k}_3} -ss'\frac{k_3(k_3-\bar{k}_3)}{(\e-1)k_0^2}.
\end{equation}
The constant $a_\ga$ follows from the normalization condition \cite{BKL5}:
\begin{equation}
\begin{split}
    \frac{|a_\ga|^2}{2} \big[1+\sum_{s'=\pm1}(|r_{s'}|^2 +\e|l_{s'}|^2)\big]&=1,\quad\text{for $\im\bar{k}_3=0$};\\
    \frac{|a_\ga|^2}{2} \big[1+\sum_{s'=\pm1}|r_{s'}|^2\big]&=1,\quad\text{for $\im\bar{k}_3>0$}.
\end{split}
\end{equation}
In the latter case, only the modes of the electromagnetic field in vacuum ($z>0$) contribute to the normalization constant since the modes describing the electromagnetic field in the medium are square integrable and their contribution to the normalization vanish for $V\rightarrow\infty$. This occurs not only in the case of the total reflection of electromagnetic waves from the crystal interface but also in the case when the medium absorbs considerably the given mode. The photonic modes \eqref{phot_modes} are characterized by the quantum numbers $\ga=(s,k_0,\spk_\perp)$ and
\begin{equation}
    \sum_\ga\equiv\sum_{s=\pm1}\int_{k_0>|\spk_\perp|}\frac{Vdk_0d\spk_\perp}{(2\pi)^3} \frac{k_0}{\sqrt{k_0^2-k_\perp^2}}.
\end{equation}
Notice that for $\e\rightarrow-\infty$ the mode functions \eqref{phot_modes} tend to the photon mode functions in the presence of an ideally conducting mirror (see, e.g., \cite{BKL5}).

In evaluating the amplitude of the surface photoelectric effect under the action of the twisted photon, it is useful to work in the basis of eigenvectors of the projection of photon spin operator onto the $z$ axis,
\begin{equation}\label{pm_basis}
    \spe_\pm=\spe_1\pm i\spe_2,\qquad\spe_3,
\end{equation}
where $\spe_i$, $i=\overline{1,3}$, are the standard basis vectors. In the basis \eqref{pm_basis}, we have
\begin{equation}\label{polarization_vects_pm}
\begin{split}
    \mathbf{f}_{s'}=&\,\frac12(\spe_+ f_-^{s'}e^{-i\vf_k} +\spe_- f_+^{s'}e^{i\vf_k}) +\spe_3 f_3^{s'},\\
    \tilde{\mathbf{f}}_{s'}=&\,\frac12(\spe_+ \tilde{f}_-^{s'}e^{-i\vf_k} +\spe_- \tilde{f}_+^{s'}e^{i\vf_k}) +\spe_3 \tilde{f}_3^{s'},\\
    \tilde{\mathbf{g}}_{s'}=&\,\frac12(\spe_+ \tilde{g}_-^{s'}e^{-i\vf_k} +\spe_- \tilde{g}_+^{s'}e^{i\vf_k}) +\spe_3 \tilde{g}_3^{s'},
\end{split}
\end{equation}
where the dependence on the azimuth angle, $\vf_k$, is explicitly shown. It follows from expression \eqref{polarization_vects} that
\begin{equation}
\begin{aligned}
    f_\pm^{s'}&=\frac{1}{\sqrt{2}k_0}(k_3 \mp s'k_0),&\qquad f_3^{s'}&=-\frac{k_\perp}{\sqrt{2}k_0},\\
    \tilde{f}_\pm^{s'}&=\frac{1}{\sqrt{2}k_0}(-k_3 \mp s'k_0),&\qquad \tilde{f}_3^{s'}&=-\frac{k_\perp}{\sqrt{2}k_0},\\
    \tilde{g}_\pm^{s'}&=\frac{1}{\sqrt{2}k_0\e^{1/2}}(-\bar{k}_3 \mp s'k_0\e^{1/2}),&\qquad \tilde{g}_3^{s'}&=-\frac{k_\perp}{\sqrt{2}k_0\e^{1/2}}.
\end{aligned}
\end{equation}
Substituting the expansions \eqref{polarization_vects_pm} into \eqref{phot_modes}, we arrive at
\begin{equation}\label{phot_mod_func_expan}
    \bs\phi_{\ga}(\spx)=e^{i\spk_\perp\spx_\perp}\big[\frac12(\spe_+\phi_{\ga-}e^{-i\vf_k} +\spe_-\phi_{\ga+}e^{i\vf_k} ) +\spe_3\phi_{\ga 3} \big],
\end{equation}
where
\begin{equation}
    \phi_{\ga\{\pm,3\}}(z)=a_\ga
    \left\{
      \begin{array}{ll}
        \sum_{s'=\pm1}r_{s'} f^{s'}_{\{\pm,3\}} e^{i k_3z} + f^s_{\{\pm,3\}} e^{-ik_3 z}, & \hbox{for $z>0$;} \\[1em]
        \sum_{s'=\pm1}l_{s'} \tilde{g}^{s'}_{\{\pm,3\}} e^{-i \bar{k}_3z}, & \hbox{for $z<0$.}
      \end{array}
    \right.
\end{equation}
In evaluating the transition amplitude of an electron under the action of a photon, the following convolutions arise:
\begin{equation}\label{phot_mod_func_contr}
    2p^\perp_{\al i} (\phi^*_{\bar{\ga}})^\perp_i=p_\perp^\al\big[\phi^*_{\bar{\ga}-} e^{i(\vf^{\bar{\ga}}_k -\vf^\al_p)} +\phi^*_{\bar{\ga}+} e^{-i(\vf^{\bar{\ga}}_k -\vf^\al_p)} \big],\qquad k^\perp_{\bar{\ga} i} (\phi^*_{\bar{\ga}})^\perp_i= \frac{k_\perp^{\bar{\ga}}}{2} (\phi^*_{\bar{\ga}-}  +\phi^*_{\bar{\ga}+}  ).
\end{equation}
These convolutions enter into the integral $w^{\bar{\ga}}_{\bar{\al}\al}$ defined in formula \eqref{wbg_baa1}.

%\newpage

\end{document}